\documentclass[12pt,preprint]{emulateapj}
\usepackage{graphicx}
\usepackage{grffile}
\usepackage{subfigure}
\usepackage{amsmath}
\usepackage[usenames]{color}

\begin{document}

\shorttitle{On the origin of chromospheric magnetic energy}
\shortauthors{Mart\'inez-Sykora et al.}
\title{On the origin of the magnetic energy in the quiet solar chromosphere}

\author{Juan Mart\'inez-Sykora \altaffilmark{1,2,3,4}  
\& Viggo H. Hansteen \altaffilmark{3,4} 
\& Boris Gudiksen \altaffilmark{3,4} 
\& Mats Carlsson \altaffilmark{3,4} 
\& Bart De Pontieu \altaffilmark{2,3,4}
\& Milan Go\v{s}i\'c \altaffilmark{1,2}}
\email{juanms@lmsal.com}
\affil{\altaffilmark{1} Bay Area Environmental Research Institute, Moffett Field, CA 94035, USA}
\affil{\altaffilmark{2} Lockheed Martin Solar and Astrophysics Laboratory, Palo Alto, CA 94304, USA}
\affil{\altaffilmark{3} Rosseland Centre for Solar Physics, University of Oslo, P.O. Box 1029 Blindern, N-0315 Oslo, Norway}
\affil{\altaffilmark{4} Institute of Theoretical Astrophysics, University of Oslo, P.O. Box 1029 Blindern, N-0315 Oslo, Norway}

\newcommand{\eg}{{\it e.g.,}} 
\newcommand{\myemail}{juanms@lmsal.com}
\newcommand{\komment}[1]{\texttt{#1}}
\newcommand{\ul}{\underline}
\newcommand{\pref}{\protect\ref}
\newcommand{\soho}{{\em SOHO{}}}
\newcommand{\sdo}{{\em SDO{}}}
\newcommand{\stereo}{{\em STEREO{}}}
\newcommand{\iris}{{\em IRIS{}}}
\newcommand{\hinode}{{\em Hinode{}}}
\newcommand{\jms}[1]{\color{black}{#1}}

\begin{abstract}
	The presence of magnetic field is crucial in the transport of energy through the solar atmosphere. Recent ground-based and space-borne observations of the quiet Sun have revealed that magnetic field accumulates at photospheric heights, via a local dynamo or from small-scale flux emergence events. However, most of this small-scale magnetic field may not expand into the chromosphere due to the entropy drop with height at the photosphere. Here we present a study that uses a high resolution 3D radiative MHD simulation of the solar atmosphere with non-grey and non-LTE radiative transfer and thermal conduction along the magnetic field to reveal that: 1) the net magnetic flux from the simulated quiet photosphere is not sufficient to maintain a chromospheric magnetic field (on average), 2) processes in the lower chromosphere, in the region dominated by magneto-acoustic shocks, are able to convert kinetic energy into magnetic energy, 3) {\jms the magnetic energy in the chromosphere increases linearly in time until the r.m.s. of the magnetic field strength saturates at roughly 4 to 30 G (horizontal average) due to conversion from kinetic energy}, 4) and that the magnetic features formed in the chromosphere are localized to this region.  
\end{abstract}


\section{Introduction}

The turbulent convection zone of the Sun continuously stretches, bends, and reconnects the magnetic field. This process converts kinetic into magnetic energy and is known as the local dynamo. The convection zone time and length scales increase with depth, so that local dynamo time-scales and growth rates also vary with depth \citep{Nordlund:2008dq,Rempel:2014sf,Kitiashvili:2015nr}. Consequently, and due to the relative short timescales at the surface compared to deeper layers, the magnetic energy growth is faster close to the solar surface. 

Studying the small-scale dynamo believed to operate in the solar convection zone is inherently difficult  \citep{Cattaneo:1999fr}. The turbulent dynamo relies on the transfer of kinetic energy in the form of turbulent motions into the magnetic energy \citep{Parker:1955rc,Childress:1995hc}. Identifying a working dynamo is a question of how efficient the amplification of the magnetic field is compared to the efficiency of the cascade of magnetic field down to the resistive scale \citep{Finn:1988jk}. {\jms Note that any work on the magnetic field will produce more magnetic energy -- thus the compression of a uniform magnetic field will locally increase the magnetic energy --. 
In local dynamo theory, one expects the magnetic energy} in a closed system, where a magnetic dynamo is at work, to change initially with a polynomial increase, followed by an exponential increase over time, before finally the magnetic energy saturates to the kinetic energy of the system \citep[e.g.,][]{Brandenburg:2005pb}. This comes about because at this point, the magnetic field is strong enough to hinder the motions of the plasma to a degree where the amplification of the magnetic field decreases. If one looks at a system that is not isolated, the transport of magnetic flux in and out of the system can modify the growth rate.

Usually the turbulent dynamo is modelled by using numerical simulations based on magnetohydrodynamics (MHD) assuming that a weak initial small-scale magnetic field is present from the onset. In reality, magnetic field is always present, still, in theory the presence of magnetic field is not strictly necessary if one takes into account Biermann's battery effect \citep{Biermann:1950yg}, i.e., that the seed of the magnetic field can be generated due to local imbalances in electron pressure \citep{Khomenko:2017sf}, but this effect is not included in MHD. 

When the magnetic energy increases exponentially, the magnetic energy is smaller than the kinetic energy of the fluid, and therefore the magnetic dynamo is in the linear regime. After the energy of magnetic field approaches the kinetic energy of the fluid, the Lorentz force is able to modify the flow of the fluid significantly, quenching the dynamo action that therefore enters a non-linear regime \citep[see the review][]{Childress:1995hc}. Non-linear dynamos are present not only in turbulent flows, but also in more organized flows, which at low Reynolds number can lead to greater growth rates \citep{Alexakis:2011zm}. For instance, the laminar flows can also lead to a non-linear dynamo process \citep{Archontis:2007kk}. These authors also found that vortices can concentrate non-linear dynamo processes.

The local dynamo in the convection zone is under vigorous debate. Radiative-MHD simulations performed by \citet{Vogler:2007yg,Rempel:2014sf,Hotta:2015nr} show the existence of a local dynamo, but the magnetic Prandtl number --- the ratio between magnetic ($Re_m$) and viscous Reynolds ($Re$) numbers ($Pr = Re_m/Re$) --- in those numerical simulations is believed to be larger than the one estimated at the solar atmosphere  \citep{Archontis:2003wq,Schekochihin:2004tg,Brandenburg:2011yk,Brandenburg:2014mu}. The efficiency of energy conversion from kinetic into magnetic decreases when the magnetic Prandtl number decreases. This dependence is different across different various types of flows and turbulence, which may suggest that the local dynamo present in the solar surface is minor. Still, high resolution observations show very confined and small-scale magnetic features within the photosphere, contrary to what is found in those simulations. Another consideration to bear in mind is that studies with varying $Pr$ numbers pay the penalty of being confined to extremely low, and perhaps unrealistic, $Re_m$ in order to achieve the small Prandtl numbers believed to characterize solar conditions \citep{Brandenburg:2011yk}. It is thus clear that further studies are required to settle the debate about the presence and nature of the local dynamo in the photosphere. 

It is of course of great interest to study how the local dynamo impacts the solar atmosphere. In the quiet Sun, there is no observational evidence of magnetic field in the chromosphere resulting from photospheric small-scale flux emergence events \citep[e.g.,][]{Martinez-Gonzalez:2009rp,Martinez-Gonzalez:2010kb}. These studies used magnetic field extrapolation from photospheric features in a few events and spectroscopic signals of the chromosphere with \ion{Ca}{2}. It is unclear whether the lack of strong observational evidence for the effects of small-scale photospheric fields on the chromosphere is caused by insufficient sensitivity, or because weak fields do not reach greater heights. In the absence of conclusive observations, models can help. Recently, \citet{Amari:2015fe} studied the impact of the local photospheric dynamo on the corona. They found that the energy flux generated by photospheric motions, including the local dynamo, is large enough to play a major role in heating the corona. However, their model does not include a detailed description of the upper solar atmosphere, the chromosphere is not well resolved and a proper radiative transfer treatment of the atmosphere is lacking. An accurate treatment of the radiative losses in the photosphere and chromosphere will impact the Poynting flux through the solar atmosphere. Here we study the impact of the photospheric quiet Sun magnetic field on the chromosphere using a state-of-the-art numerical model that does incorporate many physical processes that have previously been ignored {\jms We will show in this work how magnetic field rising into the chromosphere is by itself not able to maintain the field strength there. The consequence of this must be that an amplification of the magnetic field in the chromosphere must occur}. Our model reveals that the kinetic energy in the chromosphere can produce magnetic field in-situ. 

We first provide a short description of the radiative-MHD code and the numerical model. The results are discussed in the next section, the analysis divided between describing the magnetic and kinetic energy in the chromosphere (Section~\ref{sec:cgr}), Poynting flux (Section~\ref{sec:flx}), magnetic field structures in the chromosphere (Section~\ref{sec:fstr}) and reconnection (Section~\ref{sec:rec}). We end this manuscript with a conclusion and discussion section. 

\section{Numerical model}~\label{sec:mod}

We use a 3D radiative-MHD numerical simulation computed with the Bifrost code \citep{Gudiksen:2011qy}. The model includes radiative transfer with scattering in the photosphere and lower chromosphere \citep{Skartlien2000,Hayek:2010ac}. In the middle and upper chromosphere, radiation from specific species such as hydrogen, calcium and magnesium is computed following \citet{Carlsson:2012uq} recipes, while using optically thin radiative losses in the transition region and corona. Thermal conduction along the magnetic field, important for the energetics of the transition region and corona, is solved following the formulation of \citet{Rempel:2017zl}. 

\begin{figure}[tbh]
	\begin{center}
		\includegraphics[width=0.96\hsize]{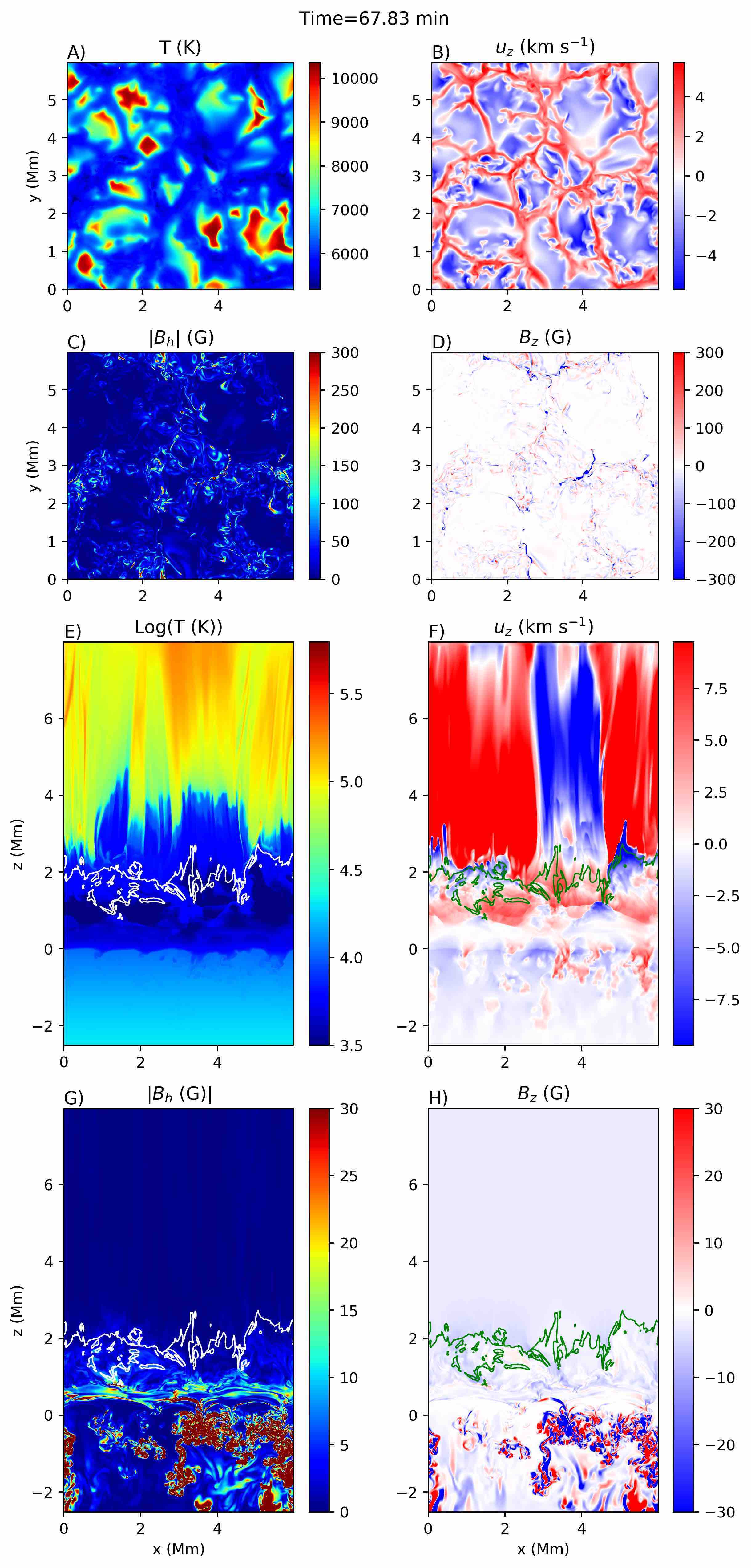}
		\caption{Horizontal maps in the photosphere (panels A-D) and vertical cuts (panels E-H) of temperature (panels A, E), vertical velocity (panels B, F), horizontal (panels C,G) and vertical (panels D,H) magnetic field show small-scale field structures within downflows in the convection zone. The numerical domain spans 6x6~Mm horizontally and covers from the upper layers of the convection zone to the lower corona. Plasma $\beta$ unity is shown with white and green contours. See corresponding Movie~1. \label{fig:cuts}}
	\end{center}
\end{figure}

Initially, the simulation only spans a vertical domain stretching from -2.5~Mm below the photosphere up to 0.7 Mm above. The photosphere is located at $z\sim0$ (where $\tau_{500}=1$). The horizontal domain spans $6\times6$~Mm in the $x$ and $y$ directions with 5 km resolution, thus the simulation includes of order $\sim 4$ granular cells along each axis. Initially, the simulation box is seeded with a uniform weak vertical magnetic field of 2.5~G. The model is run for 51 minutes, until the magnetic field complexity and strength reaches a statistically steady state with $B_{\rm rms} = 56$~G at  photospheric heights ($z=0$~Mm). 
The magnetic energy increases in the convection zone and photosphere due to the workings of the convective motion \citep[similar to that described  by][]{Vogler:2007yg,Rempel:2014sf,Cameron:2015rm}. At this point in time we extend the outer solar atmosphere up to $z=8$~Mm. The vertical $z$-axis used is non-uniform, with the smallest grid size in regions of high vertical gradients, in the photosphere, chromosphere and transition region ($dz=4$~km) and higher in the corona and in the convection zone. The density and temperature structure of the outer atmosphere was originally set by using the horizontal mean density and temperature stratification found in the pre-existing so called ``non-GOL" model described in \citet{Martinez-Sykora:2017gol} while the velocities were initialized to zero. Thanks to the strong density increase into the convection zone the transients generated by attaching an outer atmosphere do not impact deeper layers and mostly propagate outwards (see Figure~\ref{fig:cuts} and Movie~1). The total energy of these transients are also negligible compared to the steady-state situation that develops in the chromosphere (see also Section~\ref{sec:kin}). In other words, the initial stratification of the outer modeled atmosphere did not impact the steady-state solution found at later times. 

Using a potential field extrapolation, the magnetic field was computed, starting at the height where the lowest magnetic energy in the initial model was found ($z\sim 0.42$~Mm). The initial magnetic field in the chromosphere is therefore very weak (2.5~G) and nearly vertical (see Figure~\ref{fig:cuts} and Movie~1). 

The horizontal boundaries are periodic. The top and bottom boundaries are open, allowing shocks and flows to pass through. At the bottom boundary, the entropy is set so as to maintain the convective motion with an effective temperature of $\sim5750$~K. In this model we do not inject new magnetic flux but let the magnetic field exit the simulation at the top and bottom boundaries. 

\section{Results}~\label{sec:res}

We find that in this high-resolution quiet Sun model the chromospheric magnetic energy content increases substantially with time. This could in principle be caused by 1) small-scale magnetic flux emergence, 2) diffusion of magnetic field through the atmosphere, 3) and/or the magnetic energy grows in place fed by the dynamics of the chromosphere. In the remainder of this manuscript we will show that the increase of magnetic energy is 
the result of the latter scenario. 
In the following we describe the evolution of the magnetic field and kinetic energy of the plasma and their propagation through the atmosphere. In the later part of this section we will focus on the magnetic structures that appear and their evolution within the chromosphere itself and reconnection. Before describing our findings in detail we list, in Table~\ref{tab:regsim}, the heights used to separate various layers of the atmosphere. These cuts are based on properties of our simulation and not on observational properties. 

\begin{table}
	\centering
	\caption{\label{tab:regsim} Regions within the simulation
	}
	\begin{tabular}{|l|l|}
		\hline
		region/layer & height range \\ \hline  \hline
		$<\tau_{500}> = 1$ & -10~km \\ \hline
		low-photosphere & [0,0.2]~Mm  \\ \hline
		middle-photosphere & [0.2,0.45]~Mm  \\ \hline
		upper-photosphere & [0.45,0.52]~Mm \\ \hline
		low-chromosphere & [0.52,1.5]~Mm \\ \hline
		middle-chromosphere & [1.5,2.]~Mm \\ \hline
		upper-chromosphere & [2.,4.]~Mm \\ \hline
	\end{tabular}
\end{table}

\subsection{Growth of the chromospheric magnetic energy}~\label{sec:cgr}

\begin{figure}
	\includegraphics[width=0.95\hsize]{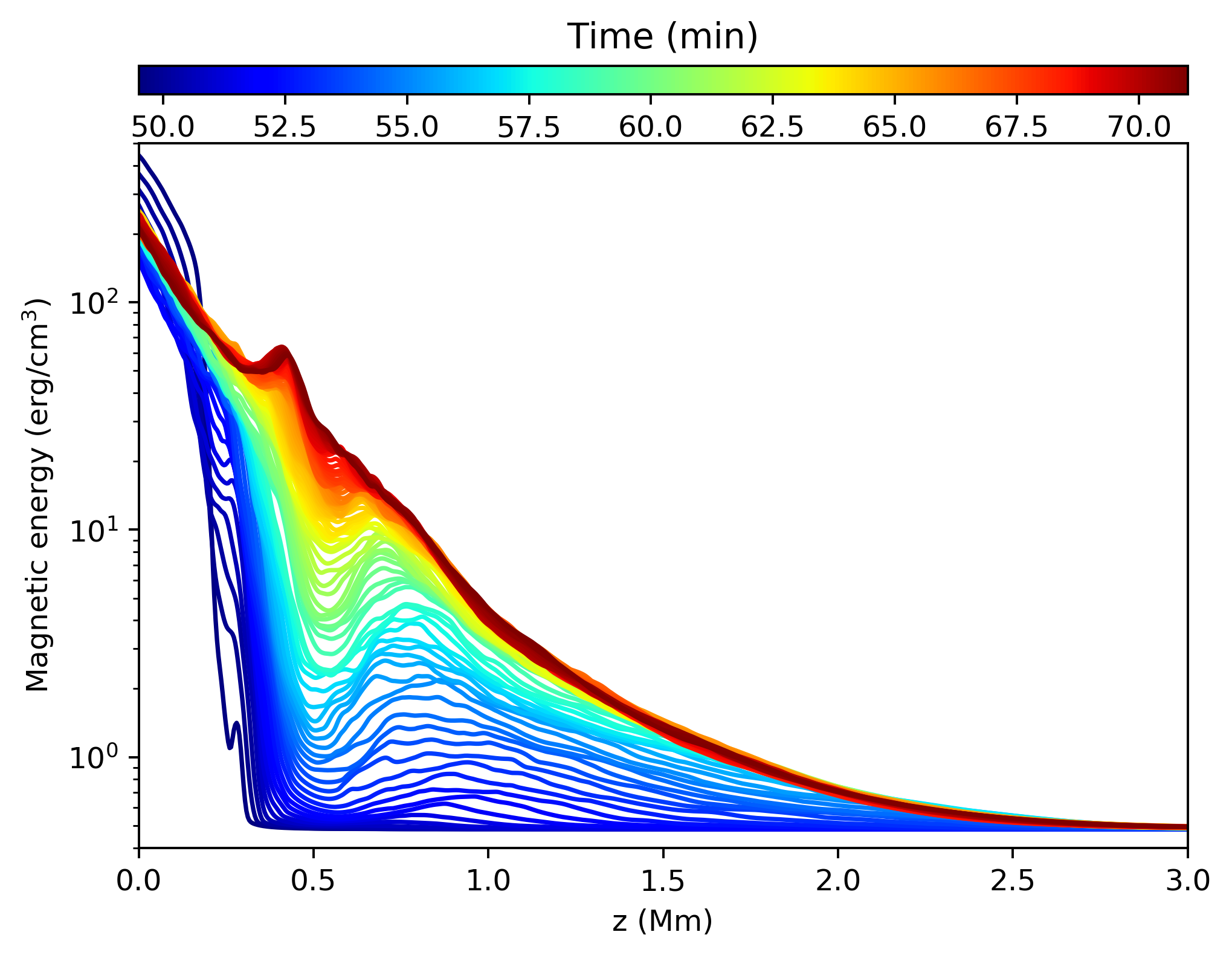}
	\includegraphics[width=0.95\hsize]{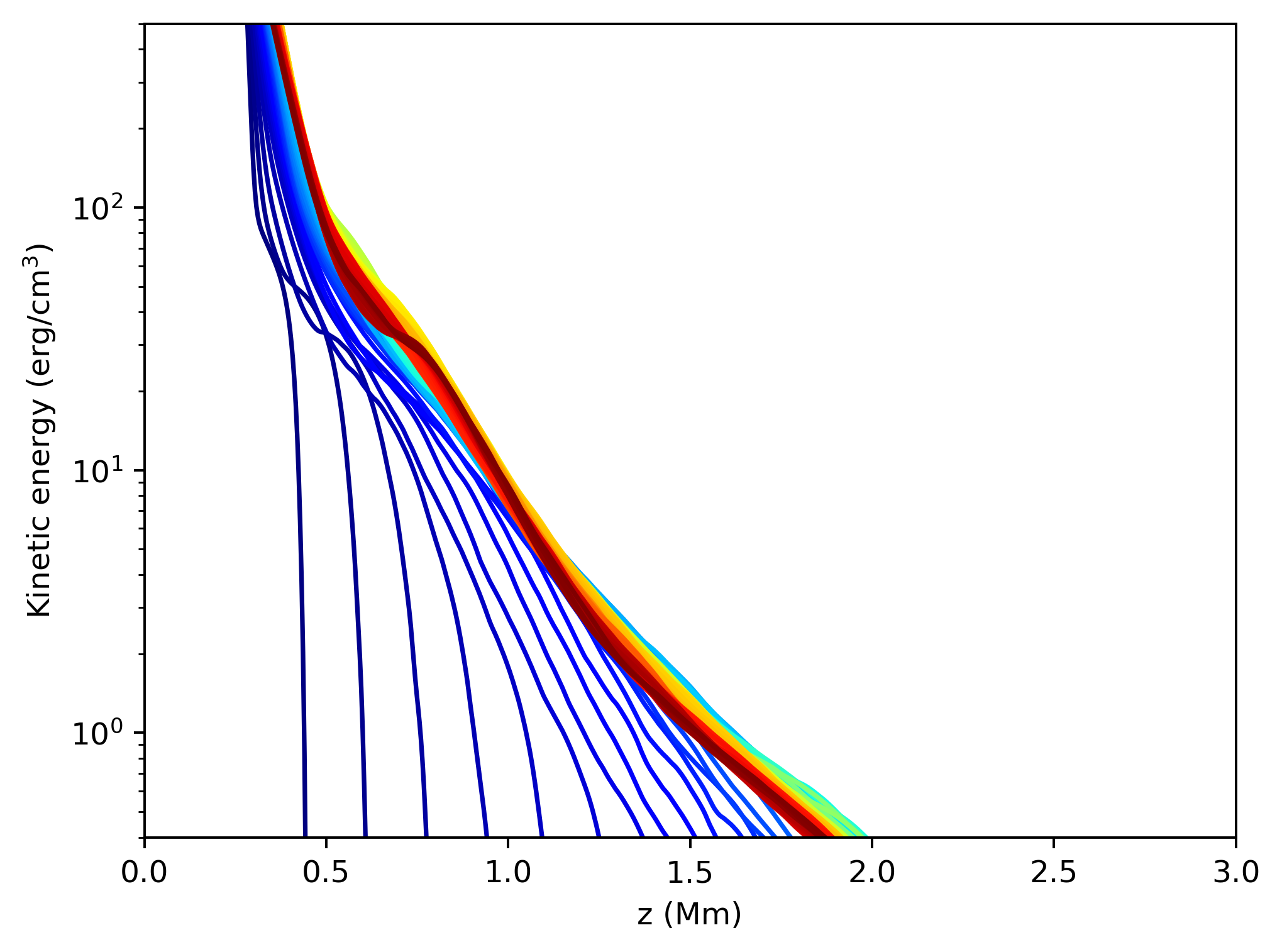}
	\includegraphics[width=0.95\hsize]{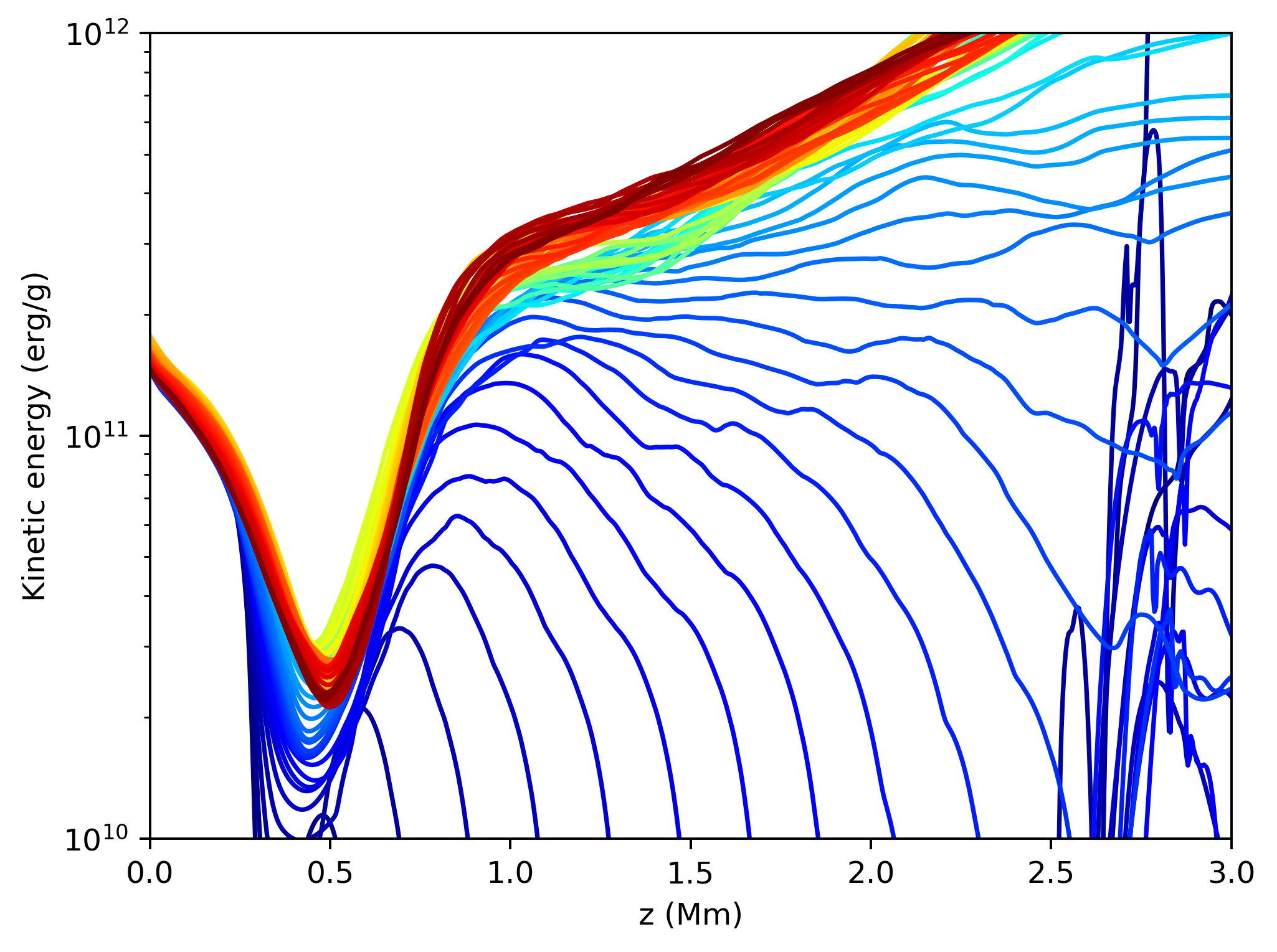}
	\caption{The kinetic energy is of the same magnitude as the magnetic energy in the low chromosphere. Horizontal mean of the magnetic energy per volume (top), kinetic energy per volume (middle) and kinetic energy per gram (bottom) are shown as a function of height (horizontal axis) and time (color-scheme). \label{fig:brms}}
\end{figure}

At the start of the simulation the kinetic energy of the chromosphere increases as acoustic waves from the photosphere propagate into the initial atmosphere. The magnetic energy in the modeled chromosphere starts to increase after only a few minutes of the initial start time ($t=50$~min\footnote{Note that, initially we started with a model that spanned only up to 0.7 Mm above the photosphere. It  took $\sim50$~min for the convective motion to build up the magnetic field in the photosphere. We then expanded the domain to include the chromosphere and corona}) of the simulation. Figure~\ref{fig:brms}
shows the mean in horizontal planes of the magnetic (top) and kinetic energies as a function of height and time. For the first fifteen minutes ($t<65$~min), the magnetic energy increases throughout the chromosphere with a local maximum centered at $z=0.8$~Mm and a local minimum near $z\approx 0.52$~Mm. At $z=0.8$~Mm, the magnetic energy first increases once the mean in horizontal planes of the magnetic energy reaches $\sim1$~erg~cm$^{-3}$. At later times ($t>65$~min), the local minimum in the magnetic energy found at $z\approx 0.52$~Mm disappears and reaches values that are at least as big as, or even greater, than that found just above. 

Previous simulations of the local dynamo in the convection zone, such as those carried out by \citet{Vogler:2007yg}, \citet{Rempel:2014sf}, or in this simulation at early times ($t<50$~min), before we added the outer chromosphere and low corona, show a sharp decrease in magnetic energy with height around $z\approx 0.3$~Mm (see blue lines in top panel of Figure~\ref{fig:brms}). This may be due to the location of the open upper boundary in these models. When the boundary is placed at greater heights we find a much higher value of the magnetic field strength at this height (yellow-red lines in the top panel of Figure~\ref{fig:brms}). 

\begin{figure}
	\includegraphics[width=0.95\hsize]{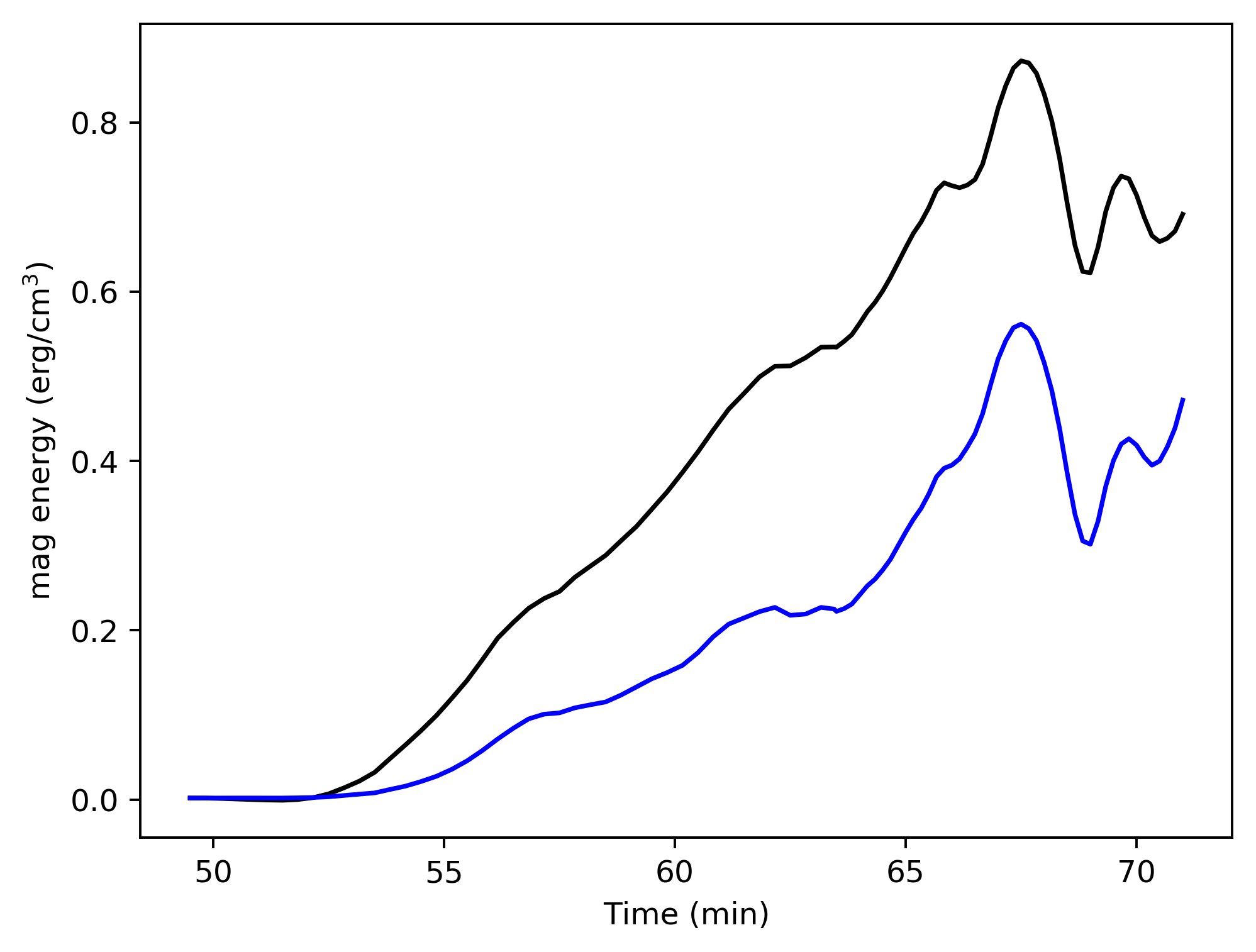}
	\caption{Total magnetic energy within the range $z=[0.52,1.5]$~Mm increases drastically in time. The magnetic flux that is connected to the boundaries of this domain have been removed by either subtracting the advected magnetic energy flux ($P_a$, resulting in the blue line for the total magnetic energy) or the full Poynting flux ($P_t$), total magnetic energy shown with a black line. See text for details. \label{fig:totbener}}
\end{figure}

We carry out our analysis by subtracting the advected magnetic flux, but also, alternately, the full Poynting flux. Figure~\ref{fig:totbener} shows, as a function of time, the mean magnetic energy within the domain $z=[0.52,1.5]$~Mm. In order quantify the magnetic energy build in-situ in the chromosphere, we have removed any in-coming or out-going magnetic flux or Poynting flux at the boundaries of this domain integrated in time, i.e., by subtracting $P_a \equiv \int_{t_0}^{t_1} B_h^2\, u_z\, \Delta t$ for the magnetic flux, or, for the Poynting flux $P_t \equiv \int_{t_0}^{t_1} [B_h\, u_z\, - B_z\,(u_x\, B_x + u_y\, B_y)] \Delta t$ at the boundaries. Where $B_x$, $B_y$, and $B_z$ are the various components of the magnetic field, $u_x$, $u_y$, and $u_z$, are the various components of the velocity field and $B_h$ is the horizontal magnetic field at the boundaries of this selected domain. The resulting evolution of the mean magnetic energy within the domain is shown with blue and black lines, respectively. Note that both methods of analysis give qualitatively the same result. In both cases, the magnetic energy increases linearly in time until reaching a steady state some 16 minutes later. Therefore, the linear increase of the magnetic energy seen in the chromosphere (blue and black lines in Figure~\ref{fig:totbener}) is not due to flux transported into the chromosphere, generated by conversion of the kinetic into magnetic energy from the convective motion, nor due to the Poynting flux through the upper and lower boundaries of the chromosphere.

The work of the Poynting flux ($B_z\,(u_x\, B_x + u_y\, B_y)$) contributes to the stretching of the magnetic field, one of the processes needed to increase the magnetic energy with the dynamics. This mechanism converts kinetic energy into magnetic energy. However, as mentioned in the introduction, the local dynamo must occur in a closed and self-contained system, i.e., no energy flux at the boundaries. Such a requirement is not achievable in the solar atmosphere, the Sun is highly connected throughout its atmosphere including the lower chromosphere. Therefore, the convection zone will disturb the magnetic field configuration in the layers above. Consequently, this can not be considered as a closed and self-contained system such as that covered by local dynamo theory. Furthermore, since the magnetic energy growth is linear and not exponential, this field magnification does not correspond to a  fast dynamo process \citep{Finn:1988jk,Cattaneo:1999fr}.

In an attempt to isolate the magnetic energy growth process in the chromosphere, as mentioned, we include the magnetic energy where we have subtracted the Poynting flux at the boundaries of this selected domain (black line in Figure~\ref{fig:totbener}). Note that in that case, the increase is even faster. This has to do with the fact that the initial vertical magnetic field is negative. In the opposite scenario, i.e., assuming that on average the vertical magnetic field sign is reversed, the Poynting flux due to horizontal motions ($ - B_z (u_x B_x+u_y B_y)$)  will become negative. In that case, the magnetic energy increase will be smaller than when subtracting the magnetic flux (blue line in Figure~\ref{fig:totbener}). 

The magnetic energy increase in the lower chromosphere comes neither from the emergence of photospheric field nor diffusion, but instead represents the increase of field in-situ. First, as mentioned above, Figure~\ref{fig:totbener} already takes into account any incoming or outgoing magnetic flux. Magnetic diffusion cannot be responsible as the magnetic energy at $z\approx 0.8$~Mm is larger than at the neighboring heights during the first 15 min. 
In fact, magnetic energy is transported away from the chromosphere while the magnetic energy in the chromosphere is still increasing (see Section~\ref{sec:flx}).

\begin{figure}
	\includegraphics[width=0.95\hsize]{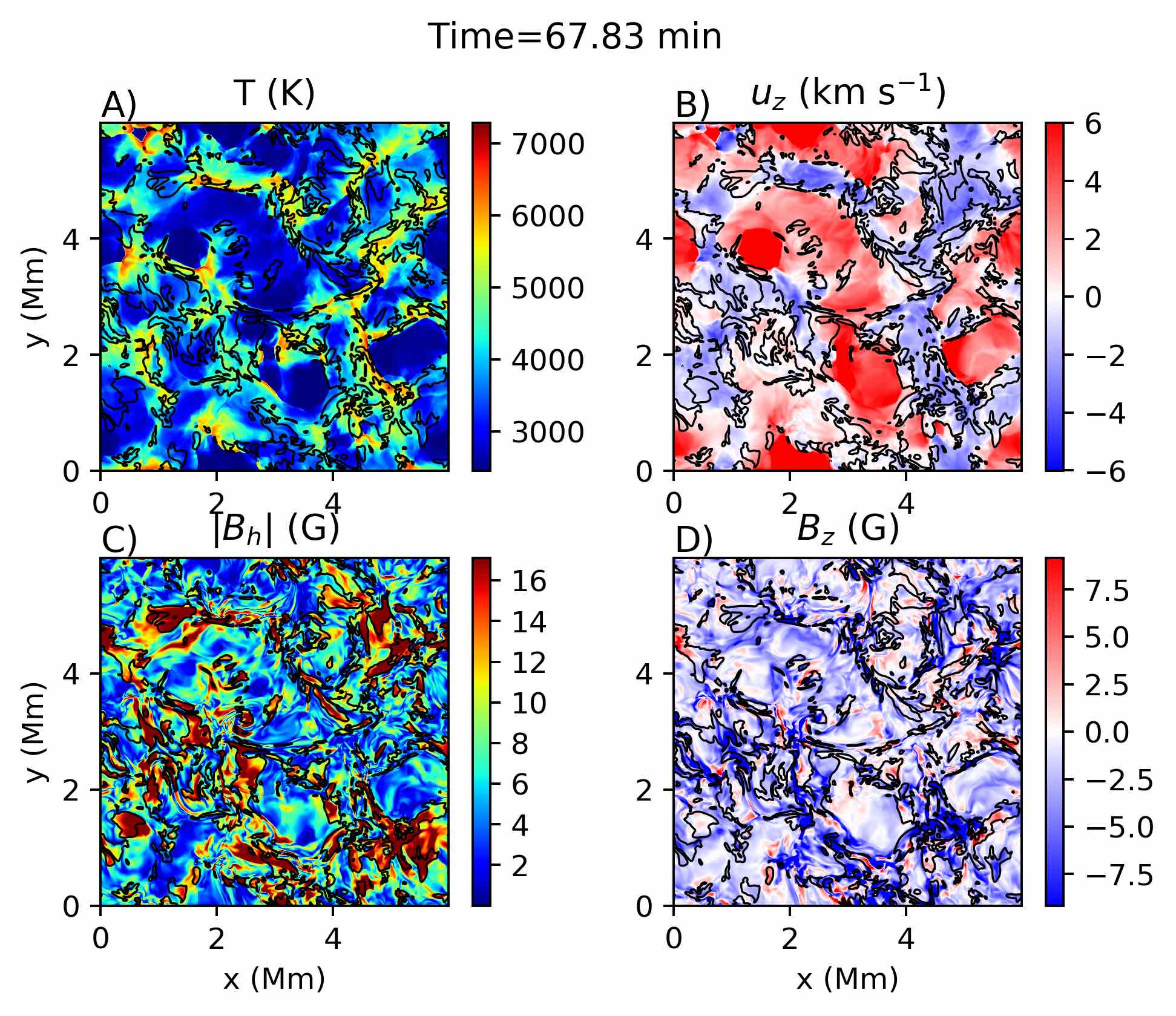}
	\caption{Temperature, vertical velocity, and horizontal and vertical magnetic field at $z=0.8$~Mm (from panels A-D, respectively) reveal that the lower chromosphere is filled with magneto-acoustic shocks. Equipartition magnetic energy is shown with black contours. See corresponding Movie~2. \label{fig:chrom}}
\end{figure}

It is interesting that the lower-chromospheric magnetic energy increases in-situ (Figures~\ref{fig:brms} and~\ref{fig:totbener}) since the lower chromosphere is not, in principle, dominated by the same physical processes as those governing the convection zone. The dynamics in the lower chromosphere are dominated by shocks driven by the convective motions occurring below. These shocks have many sources and propagate in many directions as seen in Figure~\ref{fig:chrom} and the corresponding movie. Both figure and movie show the temperature, horizontal velocity, and vertical and horizontal magnetic field maps at $z=0.8$~Mm. The equipartition between magnetic and kinetic energy are shown in black contours, i.e., locations where $E_k = E_m$. At this height, most of the time and almost everywhere, plasma $\beta$ is higher than 1, where plasma $\beta$ is the ratio of the gas pressure to the magnetic pressure. The magnetic field is compressed and advected with the shock fronts. In the movie one can also see how both the horizontal and vertical magnetic fields increase with time. 

\begin{figure}
	\includegraphics[width=0.95\hsize]{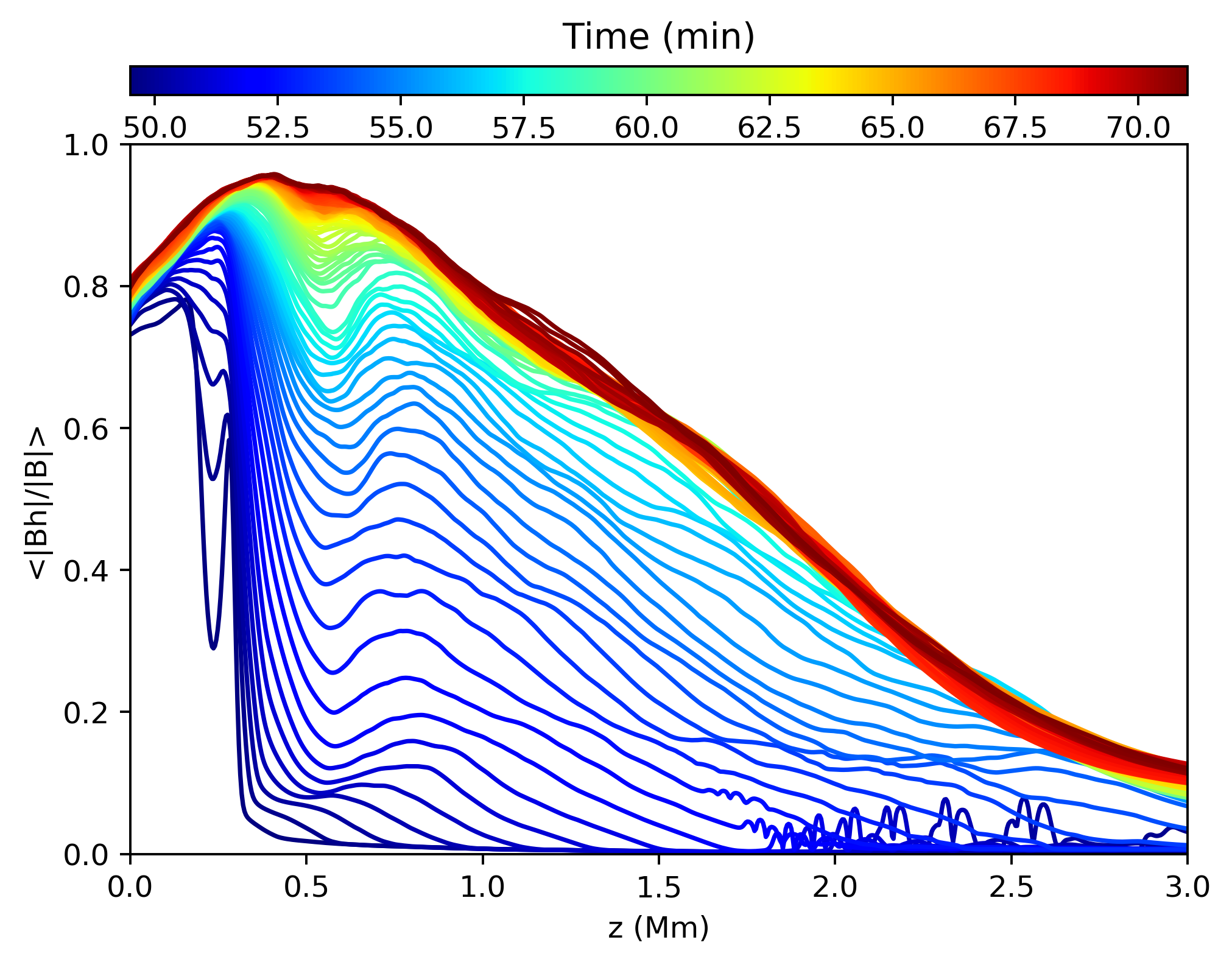}
	\caption{The magnetic field orientation becomes highly horizontal in the lower chromosphere. Horizontal mean of the ratio between the horizontal and total magnetic field strength as a function of height (horizontal axis) and time (color-scheme).\label{fig:incl}}
\end{figure}

The magnetic field orientation becomes horizontal more rapidly in the lower chromosphere than in the layers above and below. Note that initially, the model contained a weak vertical field configuration. Figure~\ref{fig:incl} illustrates the magnetic field orientation ($<|B_h|/|B|>$) as a function of height (horizontal axis) and time (color-scheme). It is interesting to see that during the first 10 to 15 minutes the field is more horizontal in the same height range within the chromosphere where the magnetic field grows faster. When the simulation reaches a steady state, the magnetic field has almost the same orientation from the upper photosphere ($z=0.3$~Mm) to the lower chromosphere ($z\sim 0.9$~Mm), i.e., it is highly horizontal in this height range.
This is evidence that the potential field extrapolation is far from the final state in which the field is largely horizontal \citep[see also][]{abbett2007}. This seems to be due to: 1) the conversion of the kinetic energy into magnetic energy which mixes the direction of the magnetic field, 2) the upper photosphere is super-adiabatic and plasma tends to expand horizontally stretching the magnetic field, 3) shocks stretch the magnetic field as shown in Section~\ref{sec:fstr}. Note that these three processes are highly connected.
As mentioned above, the magnetic energy increase in the chromosphere is not exponential, but linear with time. This is due to the fact that the lower chromosphere is not a turbulent plasma, as detailed in the following section. Note also that the simulation reaches a steady state with more or less constant magnetic field energy faster (15 minutes) than in the photosphere and much faster than in deeper layers \citep[compare with][]{Vogler:2007yg,Rempel:2014sf}.

\subsubsection{Kinetic energy}~\label{sec:kin}

The increase of the magnetic energy in the lower chromosphere shown in Figure~\ref{fig:totbener} is due to a transfer of kinetic energy. The initial chromospheric velocity field was set to zero and thus the simulation started from a static
atmosphere. Very rapidly, from the beginning of the simulation, photospheric convective motions drive a kinetic energy increase in the chromosphere. A steady state is reached in the first couple of minutes as shown in the two bottom panels of Figure~\ref{fig:brms}. The mean kinetic energy of the initial transients in the chromosphere is low enough compared to the steady state reached later in time to be excluded as a source of the increase in energy, rather that comes from the continual pumping of acoustic energy from the photosphere. Consequently, the steady state simulated atmosphere is not a result of our initial conditions but rather of the self-consistent physical processes happening in this region. 

The kinetic energy within the lower chromosphere is of the same magnitude as the magnetic energy which suggest that the kinetic energy is responsible for the growth of magnetic energy.  The fact that the kinetic and magnetic energies have similar magnitudes within the lower chromosphere in such a short period of time ($\sim 15$ minutes) means that the conversion from the kinetic energy to magnetic energy is rather efficient. 

The kinetic energy decreases more slowly with height in the lower chromosphere ($z\sim 0.8$~Mm) than it does in the upper photosphere ($z\sim 0.5$~Mm), i.e., the gradient of the kinetic energy is smaller in the lower chromosphere than in the upper photosphere. In fact, in the upper photosphere, the kinetic energy per gram (bottom panel) drops as a function of height and reaches a local minimum at $z\approx 0.5$~Mm. Above that height, the kinetic energy per gram then increases rapidly from $z=0.5$ to $z=0.8$~Mm, after the simulation has reached a steady state. In the middle and upper 
chromosphere the kinetic energy per gram keeps increasing with height which is due the density drop and flows moving along the magnetic field lines. 

\begin{figure}
	\includegraphics[width=0.95\hsize]{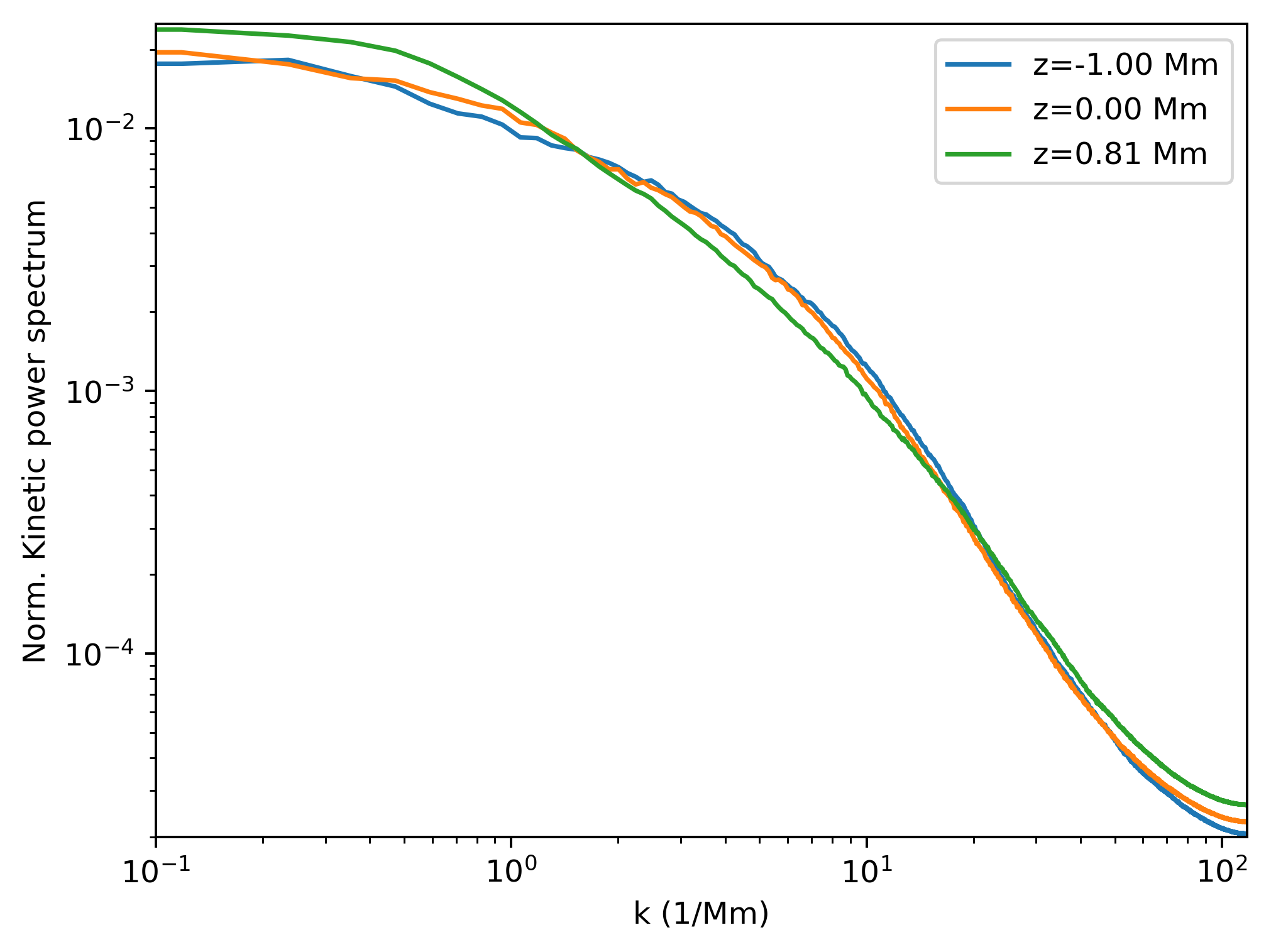}
	\caption{Normalized power spectrum of the kinetic energy is smaller at intergranular scales ($k\sim10$~Mm$^{-1}$) in the chromosphere ($z=0.81$~Mm, green) than in deeper layers such as in the  convection zone ($z=-1$~Mm, blue), photosphere ($z=0$~Mm, orange).\label{fig:kspect}}
\end{figure}	

Figure~\ref{fig:kspect} shows the normalized power spectrum of the kinetic energy at three different heights. Note that none of these follow a power law slope. Consequently, the convective motion in the photosphere, and shocks in the chromosphere are not, properly speaking, a turbulent fluid. Instead these kinematic flows have preferential scales due to the dominant processes that play a role, i.e., the processes dominating the physics and thus dynamics in each of these locations are very different; in the convection zone slow convective up-flows carry entropy upwards while faster down-flow lanes are the sites of falling cool gas. In the photosphere it is convective motions and radiative losses that dominate, while in the lower chromosphere  propagating shocks are the main source of dynamic motions.
Therefore, at intergranular length scales ($\sim 100$ km, $k\approx 10$ Mm$^{-1}$) the normalized power spectrum of the kinetic energy is greater in the convection zone ($z=-1$~Mm), and in the photosphere ($z=0$) than in the lower chromosphere ($z=0.8$~Mm). 
This may be because intergranular lane structures are not present in the lower chromosphere. The main contribution from photospheric granular scales present in the chromosphere comes from the wave patterns generated via photospheric motions (p-modes). Because of the density drop with height, these waves become shocks in the chromosphere. Since these shocks often collide, the shock pattern in the chromosphere reveals extremely thin elongated structures, which are much thinner (a few km) than the structures in the photosphere like the intergranular lanes ($\sim 100$~km).
The colliding shocks produce very narrow structures and this may explain why the relative power spectrum of the kinetic energy is greater at very small-scales in the lower chromosphere ($\sim 10$ km) than that found at lower heights. 

In the chromosphere, the magnetic field is strongest behind shock fronts or where, previously, different shock fronts have collided together, both regions containing relatively mild flows. The regions with strong magnetic field are thus not at the location of shocks nor in regions with the fastest flows. Though it is a subtle difference one may be able to see this in Movie~2. In addition, we find no correlation between the magnetic field strength and the velocity (not shown here). 

\begin{figure}
	\includegraphics[width=0.95\hsize]{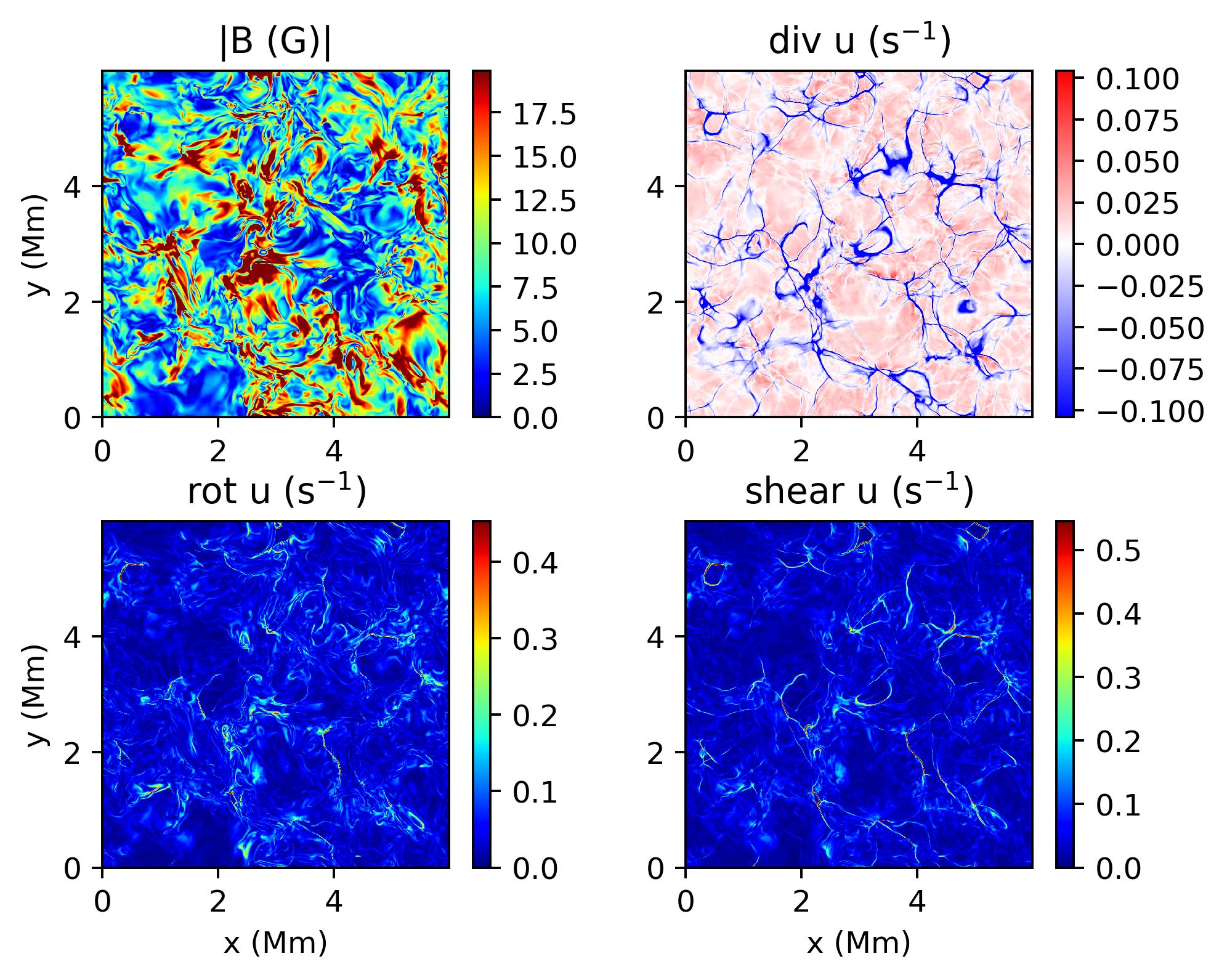}
	\caption{Magnetic field strength (panel A), divergence (panel B), rotation (panel C) and shear velocity (panel D) are shown at $z=0.8$~Mm. See associated Movie~3. \label{fig:gradmaps}}
\end{figure}

\begin{figure}
	\includegraphics[width=0.95\hsize]{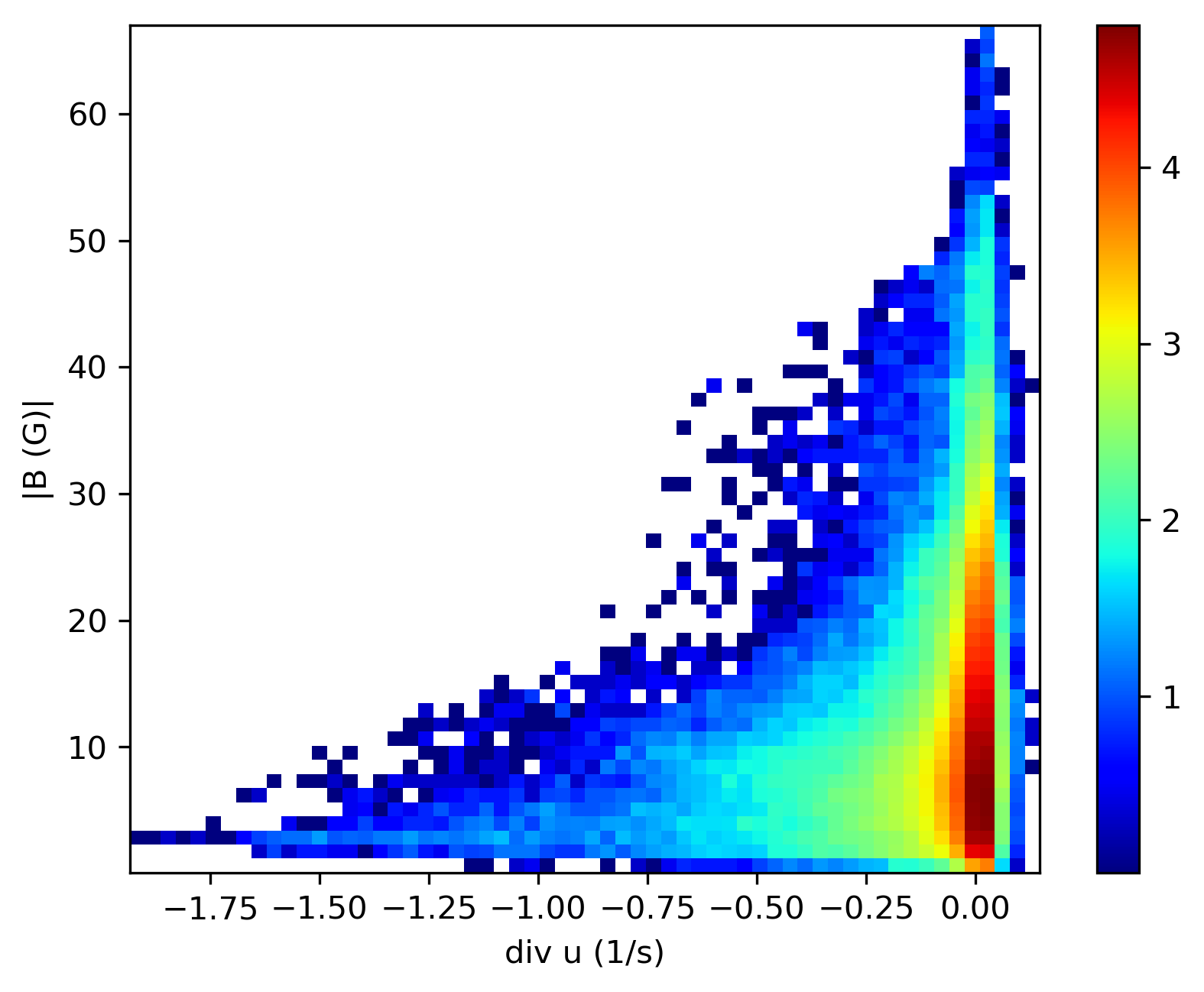}
	\includegraphics[width=0.95\hsize]{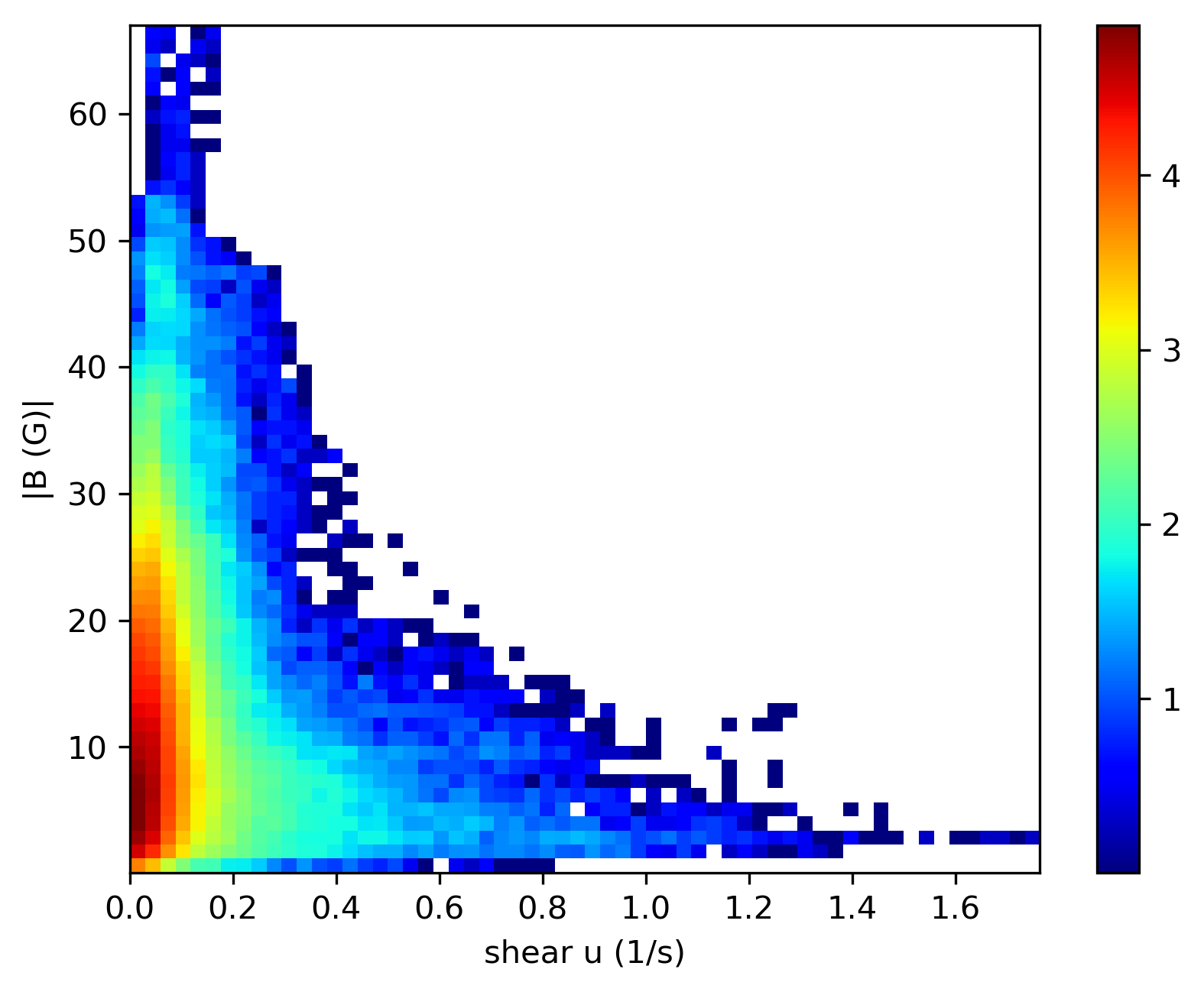}
	\includegraphics[width=0.95\hsize]{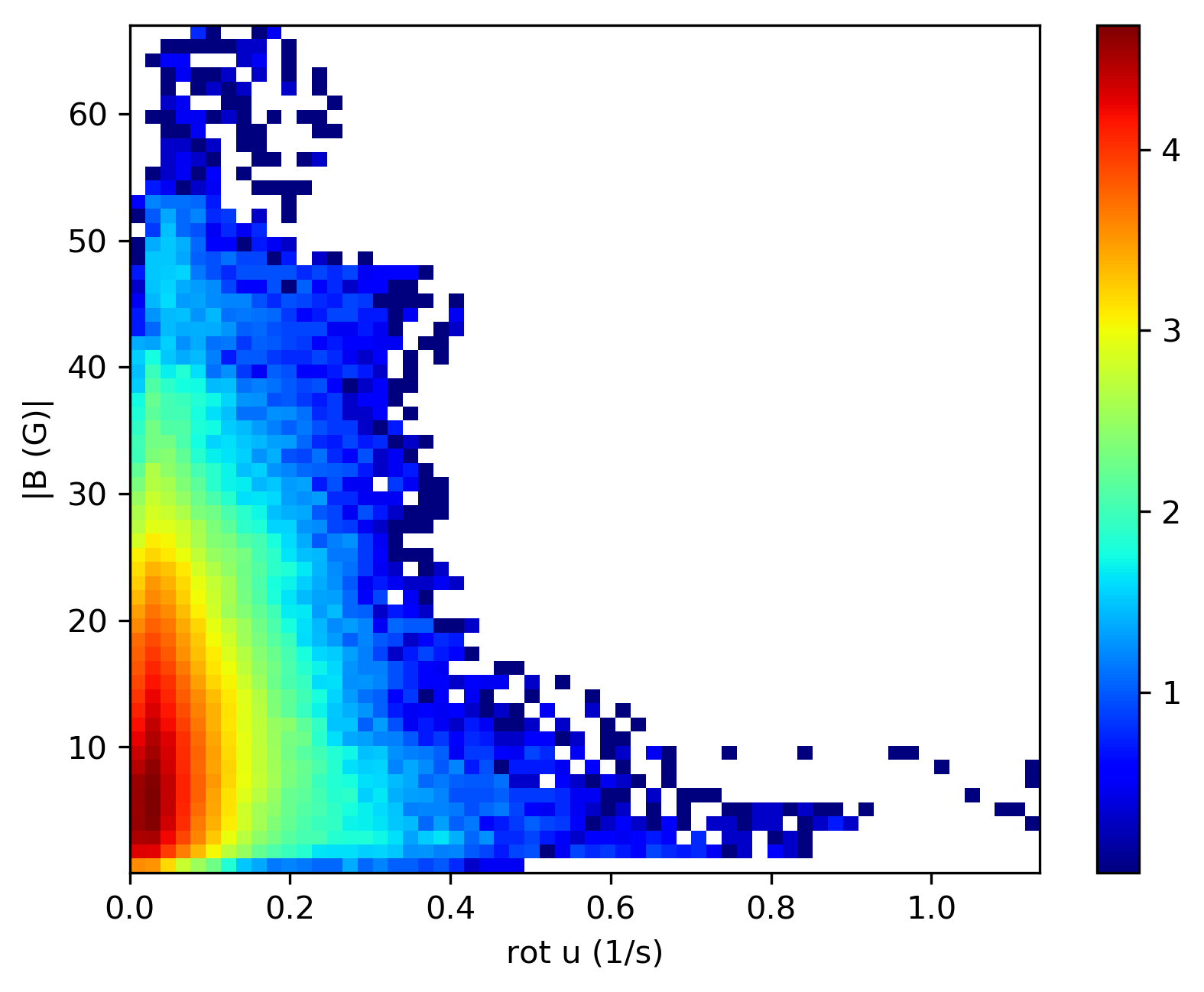}
	\caption{2D histograms of the magnetic field with divergence, shear and rotation of the velocity shown from top to bottom. \label{fig:grad2dhisto}}
\end{figure}

\begin{figure}
	\includegraphics[width=0.95\hsize]{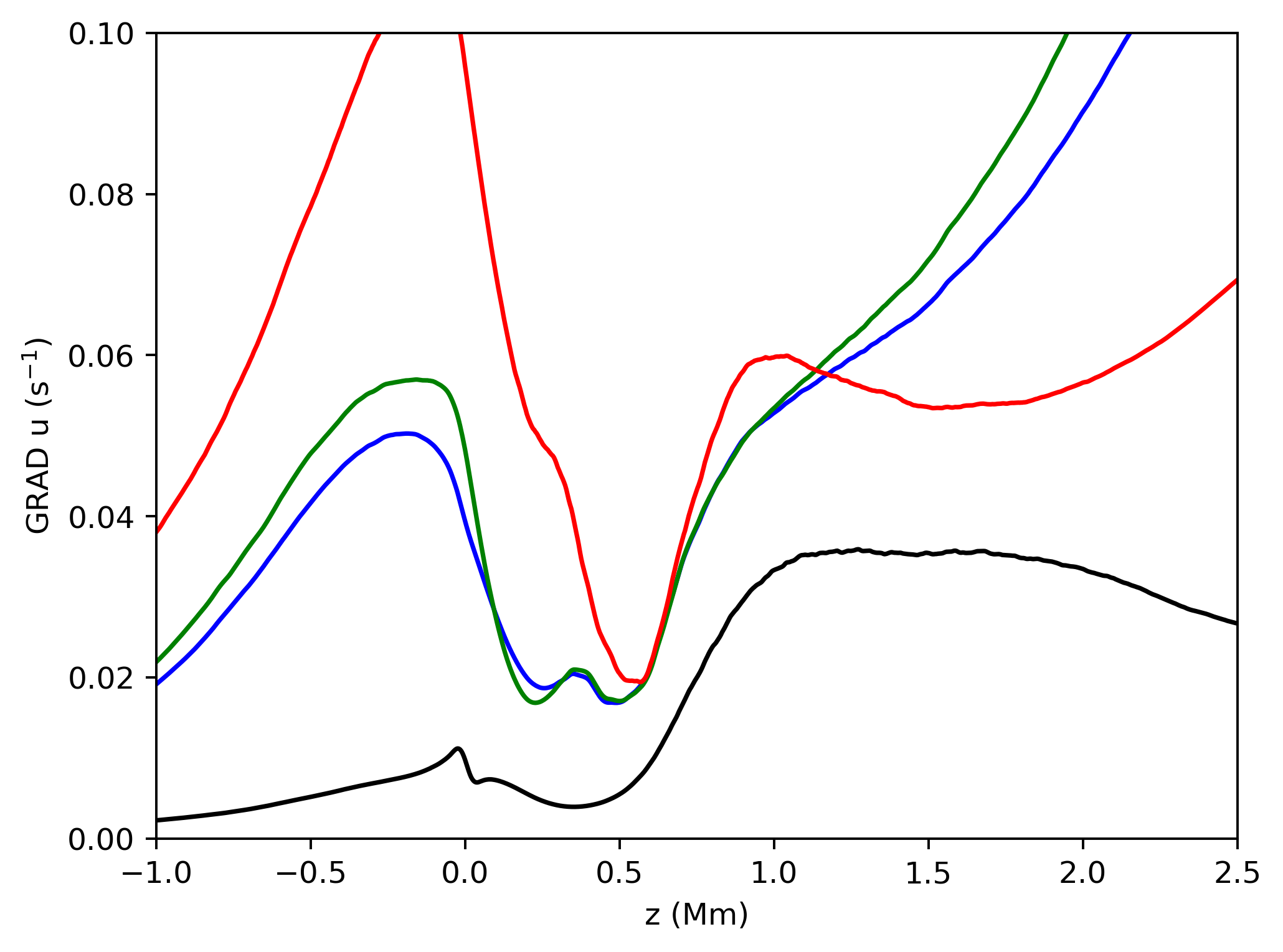}
	\caption{The various components of the gradient of the velocity increases in the lower chromosphere. Mean of the divergence (black), shear (blue) and rotation (green) as a function of height have a sharp increase in the lower chromosphere. In addition, we added the gradient of the velocity perpendicular to the magnetic field (red). 
    \label{fig:gradheight}}
\end{figure}

We also investigate shear, stretching, and compression of the plasma in the lower chromosphere. Figure~\ref{fig:gradmaps} shows maps of the magnetic field strength as well as divergence, shear and rotation of the plasma (see corresponding Movie~3). These components of the velocity gradient (panels B-D) nicely show the locations of shocks. In the shock fronts and  where the shocks collide, there is large compression, shear and rotational motions. Close visual inspection of Movie~3 shows that the magnetic field increases after a shock passes through. This is in agreement with the 2D histogram of the magnetic field with divergence, shear and rotation of the velocity shown in  Figure~\ref{fig:grad2dhisto}. Due to the fact that the magnetic field strength increases behind the shocks these 2D histograms show that regions with large magnetic field strength are in locations with low velocity gradients. It is also very important to note that the various components of the velocity gradient show a strong increase in the lower chromosphere, shown in Figure~\ref{fig:gradheight}, suggesting that these are stretching, twisting, and folding the magnetic field lines in the chromosphere. In other words, the magnetic field increases because the velocity field changes spatially. In this figure we added the gradient of the velocity perpendicular to the magnetic field (red). This term is directly linked to the process that bends and stretches the magnetic field, i.e., variations of the velocity longitudinal to the magnetic field will not bend or stretch the magnetic field.

\subsection{Poynting flux through the atmosphere}~\label{sec:flx}

Let us consider the transport of magnetic energy through  the solar atmosphere. It is well known that the Poynting flux in the convection zone is negative, i.e., magnetic field is being carried downwards from the photosphere through the convection zone \citep{Nordlund:2008dq} in quiet Sun regions. Less clear is what happens within the chromosphere and how magnetic field is transported there. \citet{Abbett:2012kc} numerical model shows that the Poynting flux changes sign in the photosphere, and becomes positive. In the upper photosphere drops to almost zero. Note that the radiative losses treatment in their model is rather simplified. 

\begin{figure}
	\includegraphics[width=0.95\hsize]{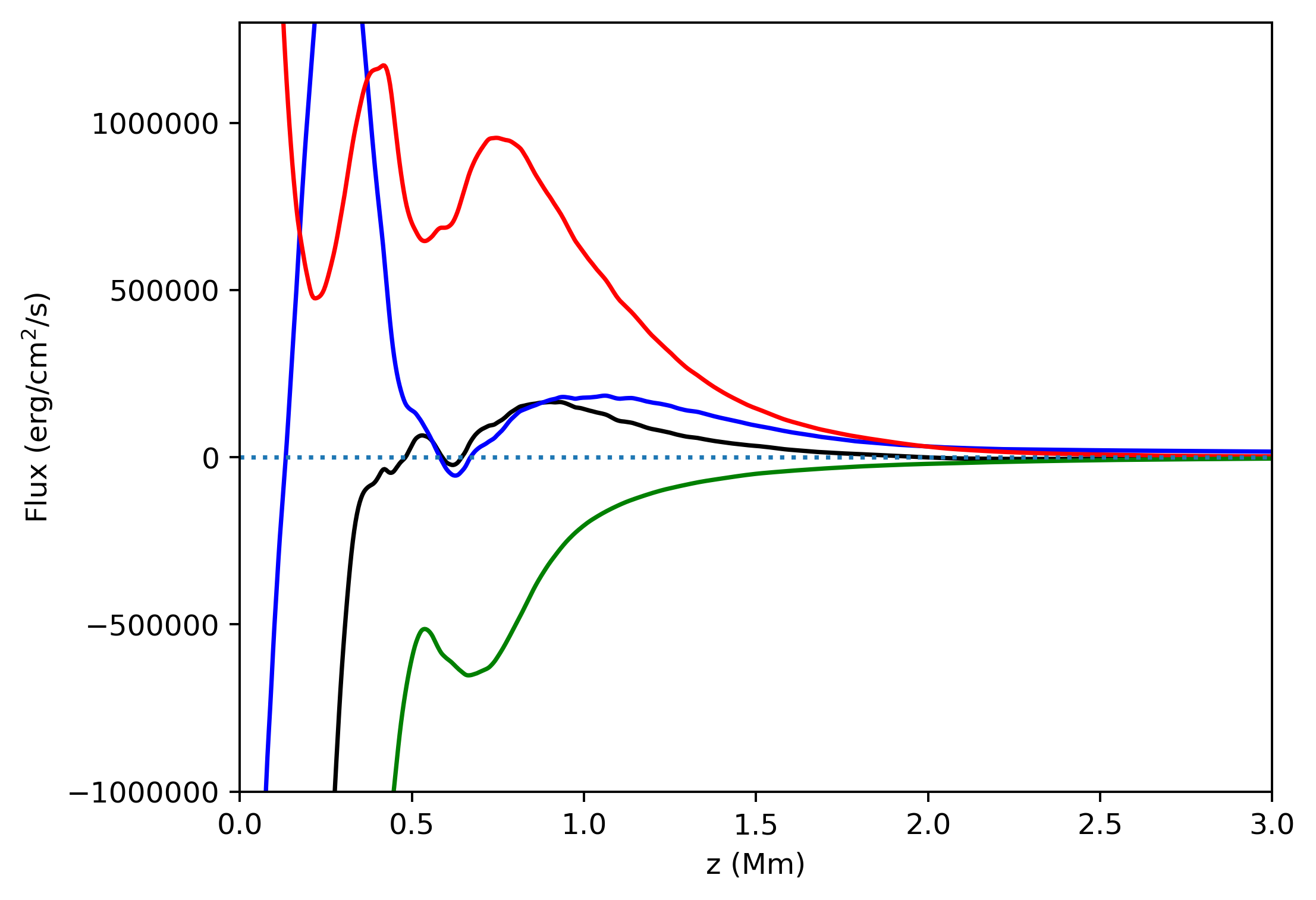}
	\caption{Horizontal mean of the advected magnetic ($P_a$, blue) and Poynting flux ($P_t$, black) as a function of height averaged in time between minute 66 and 70. In addition, we added the horizontal mean of  only positive advected magnetic flux (red) and only negative advected magnetic flux (green). Red and green lines have been divided by a factor of two (see corresponding Movie~4). \label{fig:fluxz} }
\end{figure}

Figure~\ref{fig:fluxz} 
shows the total (in black) Poynting flux  ($P_t=\int_{t_0}^{t_1} \vec{u} \times \vec{B} \times \vec{B}\, \Delta t$) and the advective component only ($P_a=\int_{t_0}^{t_1} B_h\, u_z\, \Delta t$) of the Poynting flux as a function of height, averaged over the horizontal plane and over 4 minutes in time. The shock dominated nature of the chromosphere induced by the p-mode waves from the photosphere makes the time-average necessary as these fluxes vary strongly in time. The 4 minute time average is taken from $ t= 66~$min where the chromospheric field has reached a steady state.  
This can be clearly seen in Movie~4 which shows the same lines shown in Figure~\ref{fig:fluxz} as they evolve in time. The figure shows that the advective component (blue) is on average negative up to $z\approx0.2$~Mm. Likewise, the total Poynting flux, $P_t$ is negative, also in the photosphere. 
Above this point the advective component, $P_a$ becomes positive, i.e., magnetic field is emerging, 
and increases in amplitude up to $z=0.5$~Mm. This region is where the overshoot of photospheric granular motions ends and where plasma motions become shock dominated. At greater heights, the advective component $P_a$ is negative and very close to zero within $z\approx[0.55, 0.75]$~Mm. The total Poynting flux $P_t$ in this region also goes to zero. We find that the net upper-photospheric magnetic field is not expanding to higher layers. In fact, on average a small amount of magnetic field is transported from the lower chromosphere ($z\approx[0.55,0.75]$~Mm) to deeper layers. Note that, as mentioned in the previous section, the averaged magnetic field is negative, in a more mixed polarity scenario, i.e., $<B_z> \approx 0$, the total Poynting flux as a function of height (black line) will be dominated by advection of magnetic field, and closer to the blue line. 

In Movie~4, the advected magnetic flux ($P_a$) within $z\approx[0.55,0.75]$~Mm is seen to increase from relatively large negative values (at $t\approx~56$) to nearly zero towards the end of the simulation. Consequently, during this period, at heights of $z\approx0.5$~Mm the magnetic field grows due to advected flux coming from both above and below. Above $z=0.75$~Mm, the 
$P_a$ is positive and  the magnetic field expands towards the corona.

By comparing Figure~\ref{fig:totbener} and Figure~\ref{fig:fluxz} one can see how some of the magnetic energy generated within the lower chromosphere is both expelled to lower layers ($z\sim 0.5$~Mm) and 
transported to greater heights ($z > 0.8$~Mm). This confirms that the lower chromosphere is continuously converting kinetic energy into magnetic energy. 

In the simulation the net advective component of the Poynting flux,  $P_a$ is downwards in the upper photosphere due to the entropy drop and the turnover of the convective motions. In the photosphere, the plasma is super-adiabatic, consequently, most of the plasma is advected (together with the magnetic field) to the downflows. Therefore, the magnetic field must be quite buoyant to cross this region \citep{Acheson:1979lr,archontis2004}. 
Despite this, a small fraction of magnetic field does reach the chromosphere (as seen in the red line Figure~\ref{fig:fluxz}), and there are emerging events that reach the chromosphere. However, the net transport of magnetic field is downward.  
(Investigating the dynamics and evolution of the photospheric magnetic field that does reach the chromosphere requires a separate study and is outside of the scope of this work. This component of the field is not important in the simulation described here). 

\subsection{Field structure}~\label{sec:fstr}

An important aspect of this study lies in revealing the magnetic field
structure and its evolution within the lower atmosphere. We focus on how chromospheric dynamics are able to stretch, twist, fold, and reconnect the magnetic field. In this section we will describe the different types of magnetic structures formed in-situ in the chromosphere as a result of these actions. 

\subsubsection{Photospheric field structures}~\label{sec:phstr}

\begin{figure*}
	\includegraphics[width=0.32\hsize]{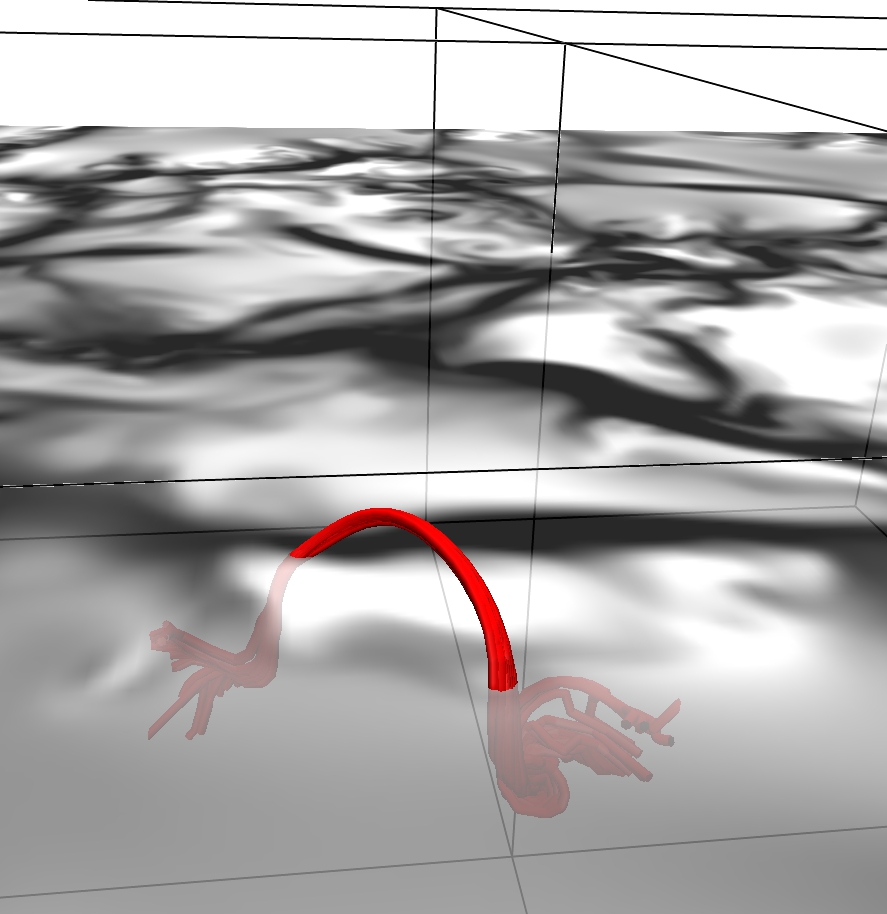}	
	\includegraphics[width=0.32\hsize]{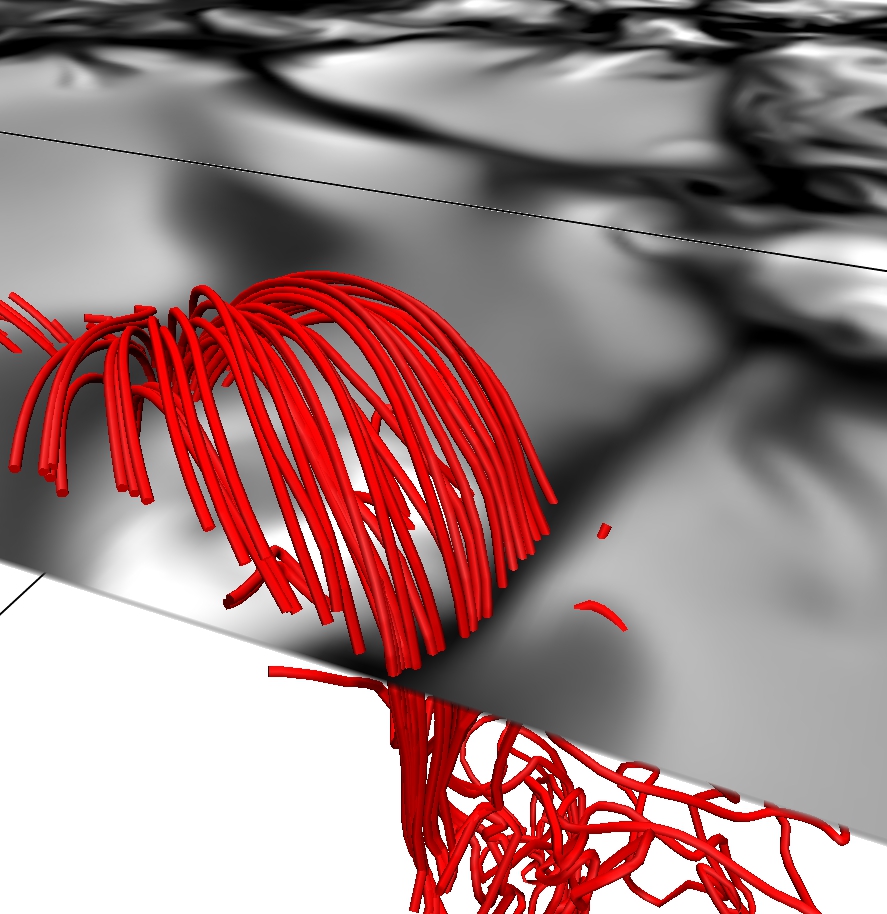}
	\includegraphics[width=0.32\hsize]{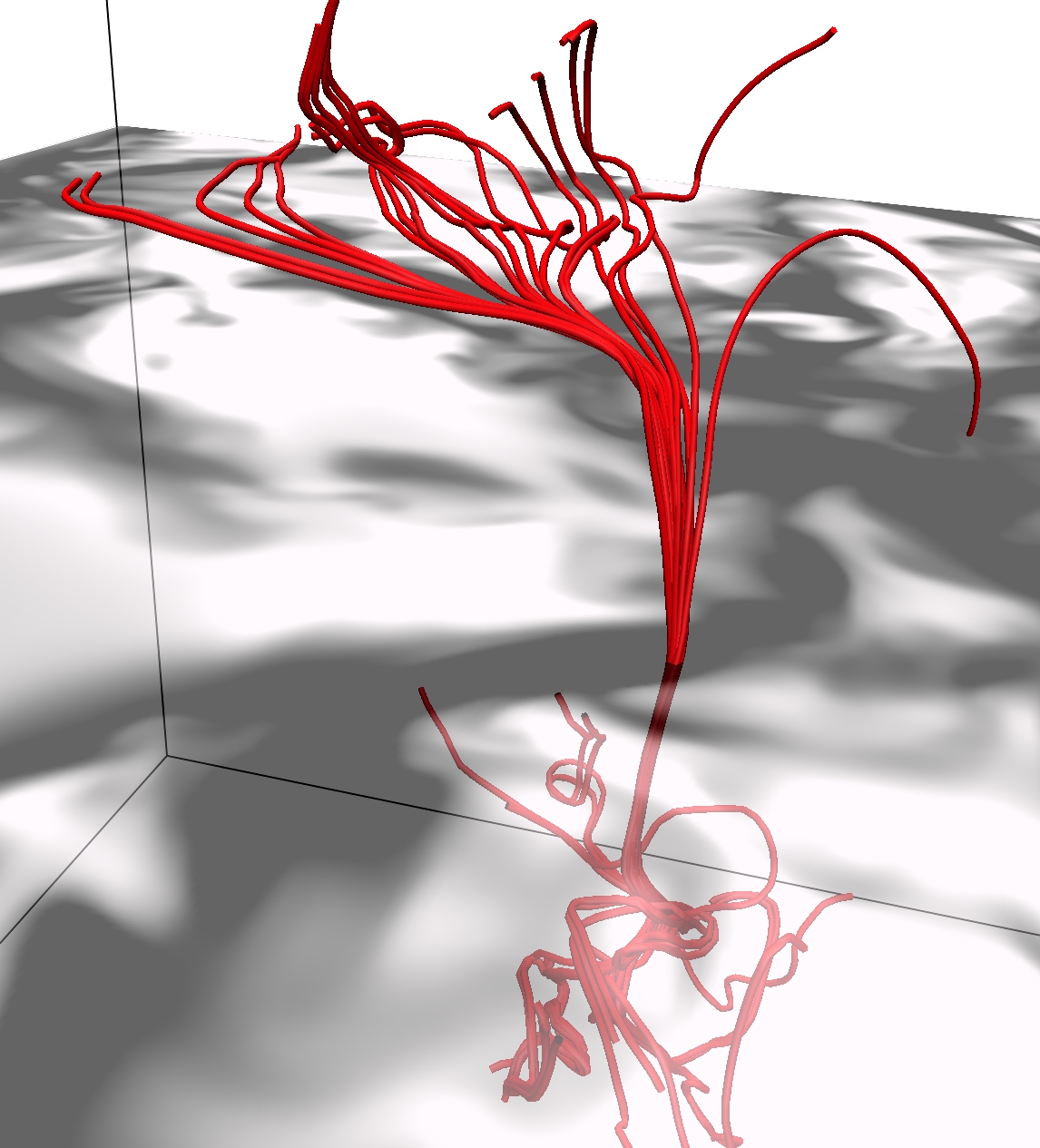}
	\caption{The photosphere is filled with emerging flux tubes (left) and flux tube canopies (right) and we also find a few cases of flux sheets (middle). The horizontal map is vertical velocity cut at the photosphere within $2$~km~s$^{-1}$ (white) and $-2$~km~s$^{-1}$ (black). \label{fig:tbandsheets}}
\end{figure*}

Before we describe the structures in the chromosphere let us for comparison give a very brief description of the structures found in the photosphere. We find that our modeled photosphere is filled with thin emerging flux-tube loops that cover only portions of the granules, 
like the example shown in the left panel of Figure~\ref{fig:tbandsheets}. We also find examples of weak flux-sheets that cover entire granular cells, e.g., as shown in the middle panel of Figure~\ref{fig:tbandsheets}. Such structures have been analyzed in depth by \citet{Moreno-Insertis:2018hl} and observed by \citet{De-Pontieu:2002by} and \citet{Centeno:2017mi}. An example of a third type of field structure that often appears in the photosphere is the magnetic field canopy that expands from the photosphere to greater heights 
\citep[seen in the right panel of Figure~\ref{fig:tbandsheets}, see also  ][]{Sainz-Dalda:2012qf,de-la-Cruz-Rodriguez:2013bs}. 
Since this numerical model has very weak large-scale connectivity, the magnetic field in the canopies is rooted in intergranular lanes, and spreads drastically in the upper photosphere and lower chromosphere instead of reaching all the way to the corona. The example shown in Figure~\ref{fig:tbandsheets} reaches a height of $z=0.5$~Mm. Aside from spatial resolution, the main differences between the model shown here and the two models described in \citet{Moreno-Insertis:2018hl} and \citet{Sainz-Dalda:2012qf} are that the model in the current  paper does not include magnetic flux injected through the bottom boundary, and does not have an initially strong magnetic field configuration. Instead, the magnetic field has been built up in the photosphere through the action of kinetic into magnetic energy conversion from the convective motions starting from a very weak vertical field. In addition, due to the high spatial resolution and resulting small-scale structures found in the downflows, we find even thinner flux tubes than in the previous models. Another difference is that many of the flux sheets and tubes in our model have highly curved and reconnected field, i.e., the magnetic field lines have a highly complex connectivity within the photosphere. 

\subsubsection{Chromospheric field structures}~\label{sec:chstr}

Consider now the chromosphere, the first thing noticed is that the magnetic field structure there is very different compared to the photospheric structures described above. The magnetic field connectivity between photospheric and chromospheric structures is very complex and it is difficult to directly trace \citep[see also][]{abbett2007}. Chromospheric magnetic features remain in the chromosphere without any clear connection to the photosphere at all, a field line typically covers distances of many megameters at chromospheric heights before descending (or ascending) and connecting to the photospheric (or coronal) field. 

\begin{figure*}
	\includegraphics[width=0.49\hsize]{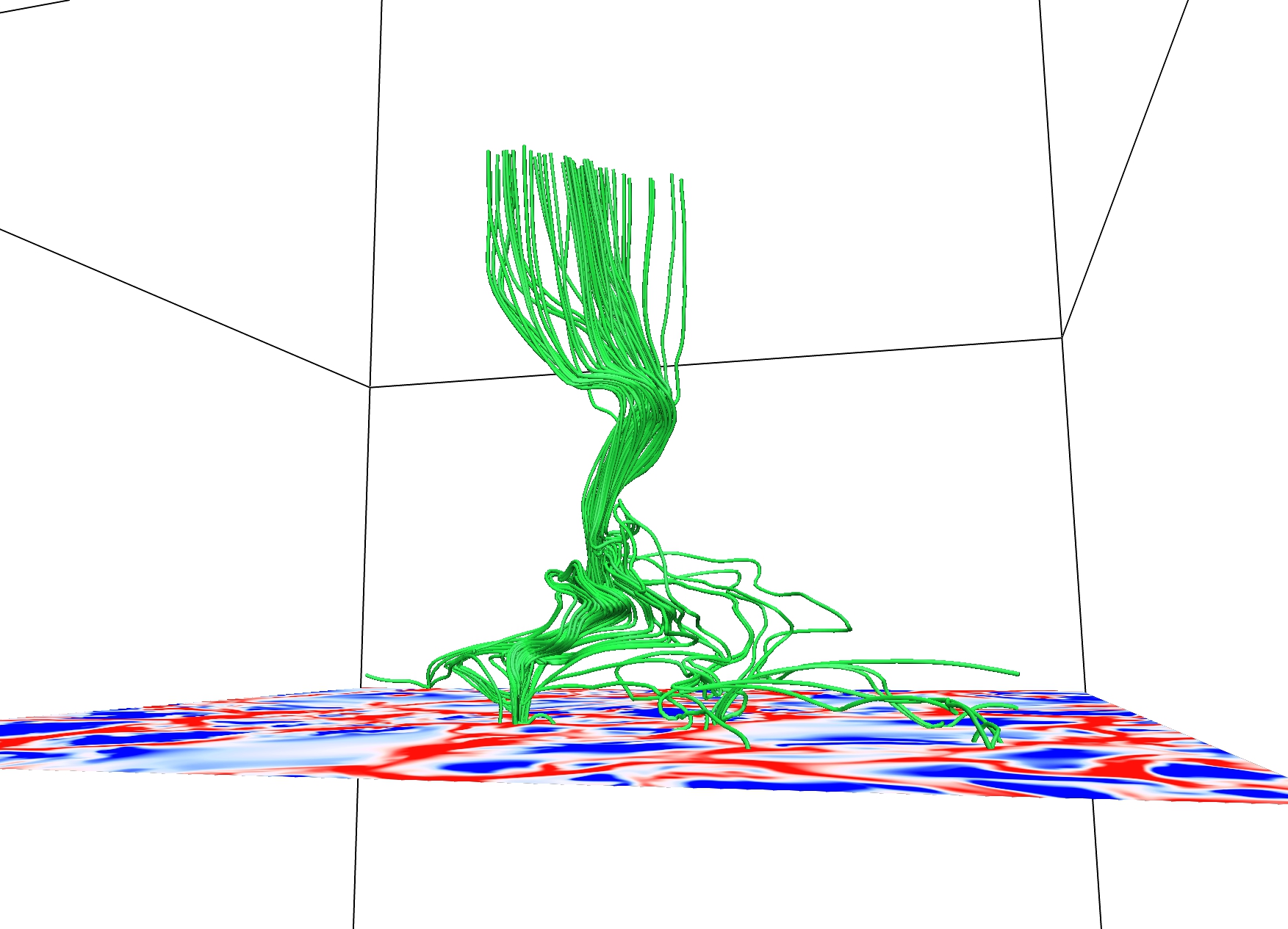}
	\includegraphics[width=0.49\hsize]{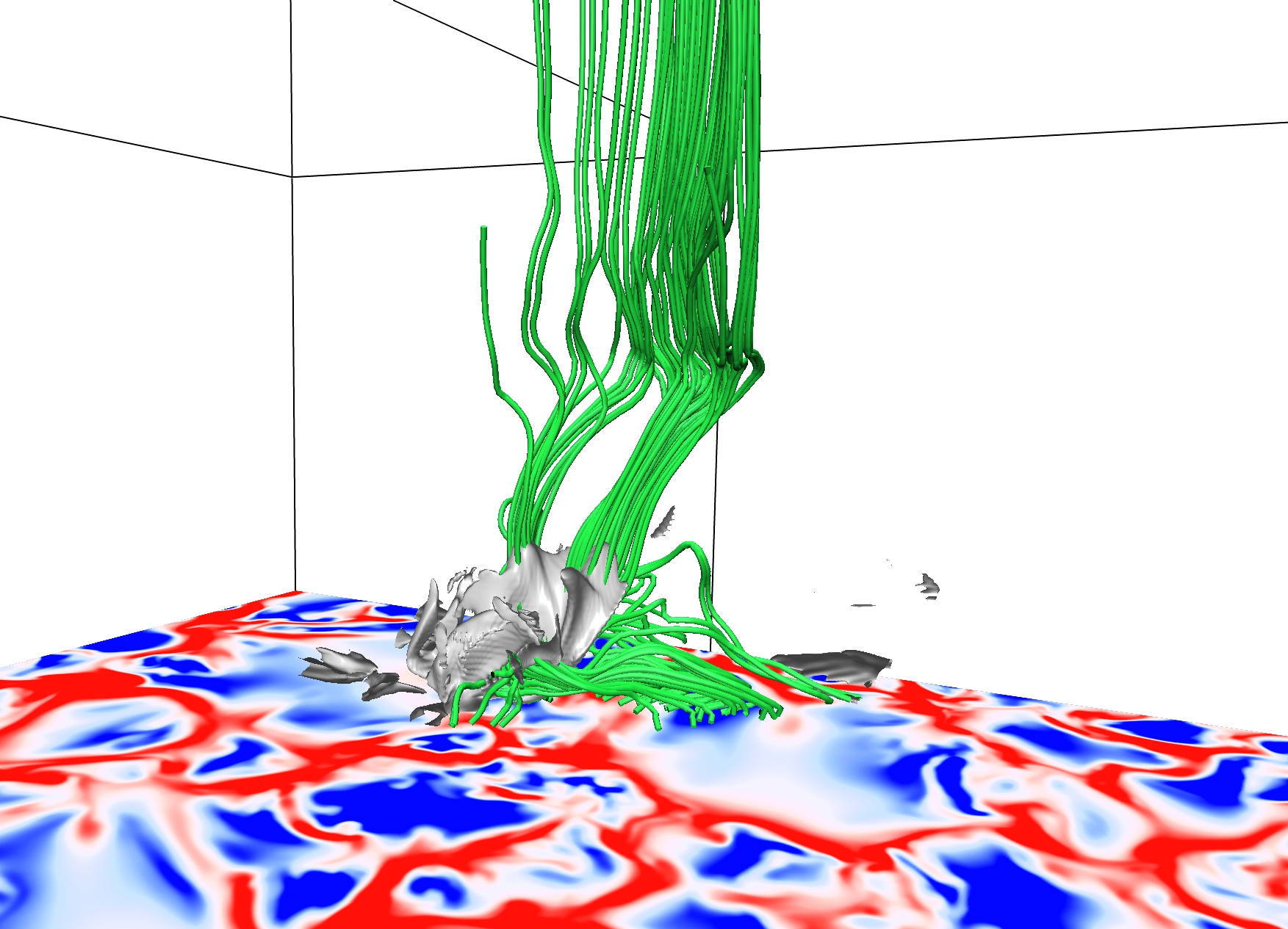}
	\includegraphics[width=0.49\hsize]{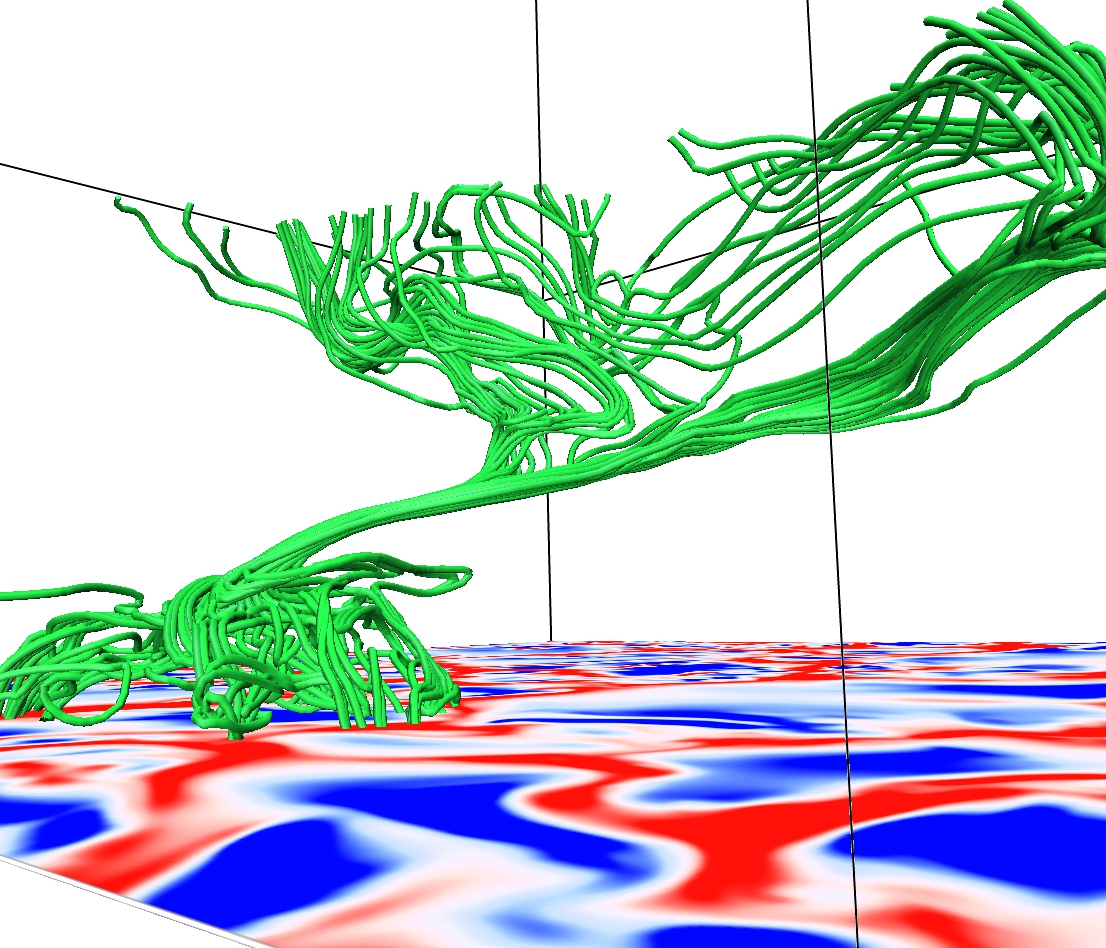}
	\includegraphics[width=0.49\hsize]{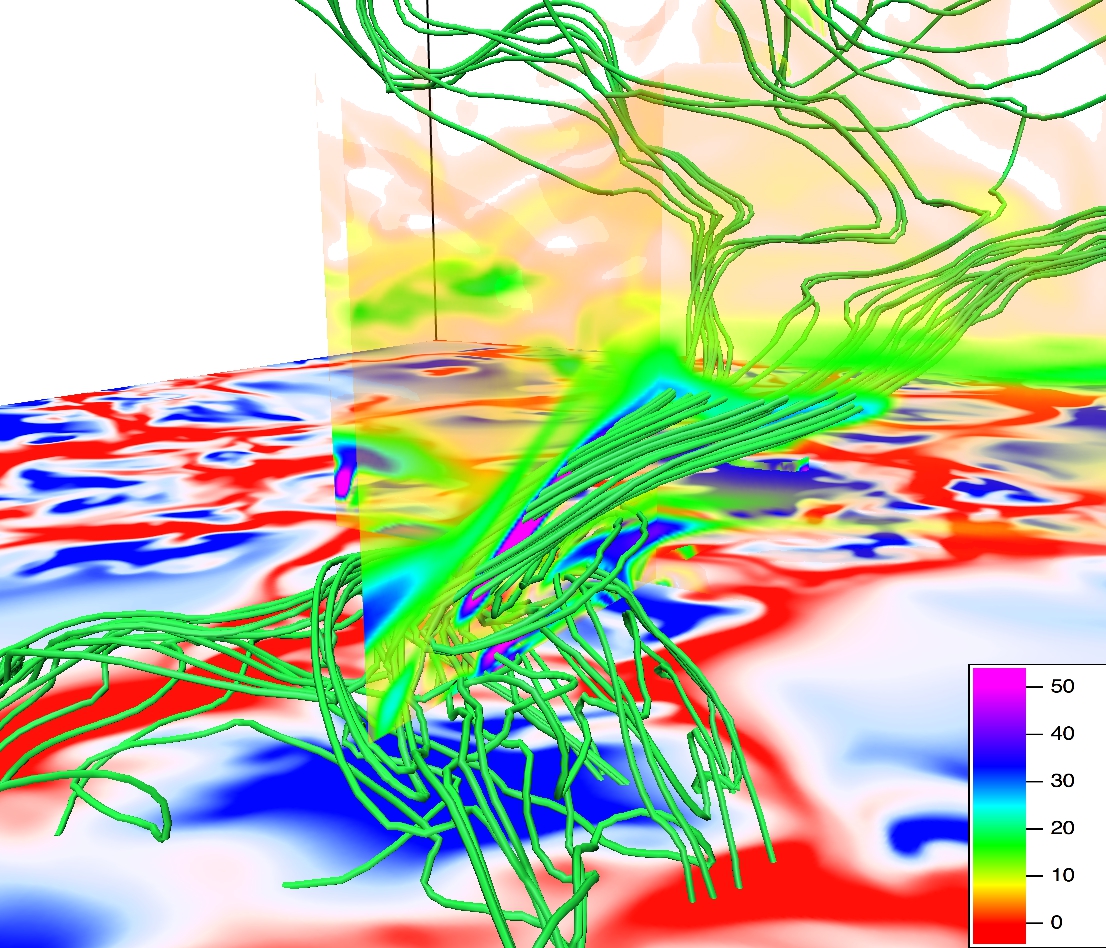}
	\caption{We found two types of magnetic field structures in the lower chromosphere. Top panels show vertical flux tube concentrations formed at the chromospheric shock fronts. The bottom panels show flux sheets of up to 64~G. Movie~5 corresponds to the bottom panels. The blue-white-red horizontal cut shows $\pm2$~km~s$^{-1}$ vertical velocity at $z=0$. The magnetic field strength at 30 G is shown with the grey isosurface in the top-right panel and the two vertical rainbow probes in the bottom-right panel show the field strength in Gauss.\label{fig:chst}}
\end{figure*}

Let us now turn to the structures found in the lower chromosphere. Two types of structures are found to dominate the very complex field configurations associated with the local growth of chromospheric magnetic energy. Both cases share some properties: at one or both ends, the magnetic field lines defining the structures expand like the petals of a flower, i.e., magnetic field lines spread out in a plane or cup shaped envelope. One or both ends of the structure can go to greater or lower heights. Both types of structure are formed in-situ in the chromosphere. Finally, they have in common that they are very confined along, at least, one axis, i.e., the structures appear as thin (elongated or short) sheets or thin flux tubes. The widths can be as small as a few tens of kilometers. 

The first type of structure, usually stronger and reaching greater heights within the chromosphere than the other, is oriented in the vertical direction with field lines localized as in a flux tube. They are usually found behind shock fronts or where different shock fronts have collided. One example can be seen in the top panels of Figure~\ref{fig:chst} and Movie~5. 
The structure shown reaches field strengths of up to $\sim 65$~G and heights of up to $1.5$~Mm before weakening due to expansion. The structures of this type that we have managed to follow have lifetimes from a few tens of seconds up to a few minutes and travel horizontally through the chromosphere at apparent speeds between 0 and $4$~km~s$^{-1}$. In order to measure the apparent horizontal velocities of the vertical magnetic field structures visible at the chromospheric heights, we applied Local Correlation Tracking \citep[LCT\footnote{LCT is a technique that correlates small regions in two consecutive images to determine displacement vectors. Those small regions are defined by Gaussian tracking window whose full-width at half-maximum (FWHM) is large enough to contain significant structures to be tracked. Here, we use a tracking window with a FWHM=0.6~Mm. To make a smooth transition between images we also used moving-average over three consecutive frames whose cadence was 10 seconds.};][]{November:1988oj}. 
The apparent speeds of the magnetic field structures are roughly similar to the real plasma flows. 

The second type of structure is a thin sheet, lying almost parallel to the surface in the lower chromosphere ($z\approx 0.6-0.8$~Mm) as shown in the bottom row of Figure~\ref{fig:chst} and Movie~6. Note that these sheets are very different from the photospheric flux sheets described at the beginning of this section \citep{Moreno-Insertis:2018hl}. These structures form in-situ and not due to flux emergence and compression in the overshoot layer. In addition, the structure's scale size and shape are not associated with granular cells. These structures are flat with horizontal widths of several hundred kilometers, occurring over a narrow height range (a few tens of km), and as long as a couple of granular cells. The examples of such sheets that we have analyzed are seen to be created as a consequence of interactions between the tube like structure described previously with horizontally traveling shocks. In this case the magnetic field is strongly stretched deforming tubes into sheet like structures. These can live up to $\sim15$ minutes. They weaken towards the end of their lifetime due to reconnection and plasma expansion. The sheet like structures can travel at horizontal speeds $\le 2$~km~s$^{-1}$ and can also be moved to deeper layers, where reconnection more easily occurs with photospheric field lines. The example shown in the lower panels of Figure~\ref{fig:chst} reaches magnetic field strengths of up to $65$~G, and towards the end of its life is found between $z=0.5$ and $0.65$~Mm above the surface. As in the first case, the connectivity to the photosphere is broadly spread, or the structure might not be properly connected to the lower layers at all. Consequently, both ends of the field structure spread over many different locations. In this case in particular, one end is in the photosphere, straddling a granule while the other end spreads over a wide region in the corona. 

The flux tubes described above are formed by passing 
shock fronts or by collisions between shocks. Consequently, they are not connected to any single location in the photosphere, instead spreading over a large area, into many and complex photosperic flux tubes and sheets. Sometimes, these structures travel in a different direction than the patterns in the photosphere below them. 

\subsection{Reconnection}~\label{sec:rec}

\begin{figure*}
	\includegraphics[width=0.24\hsize]{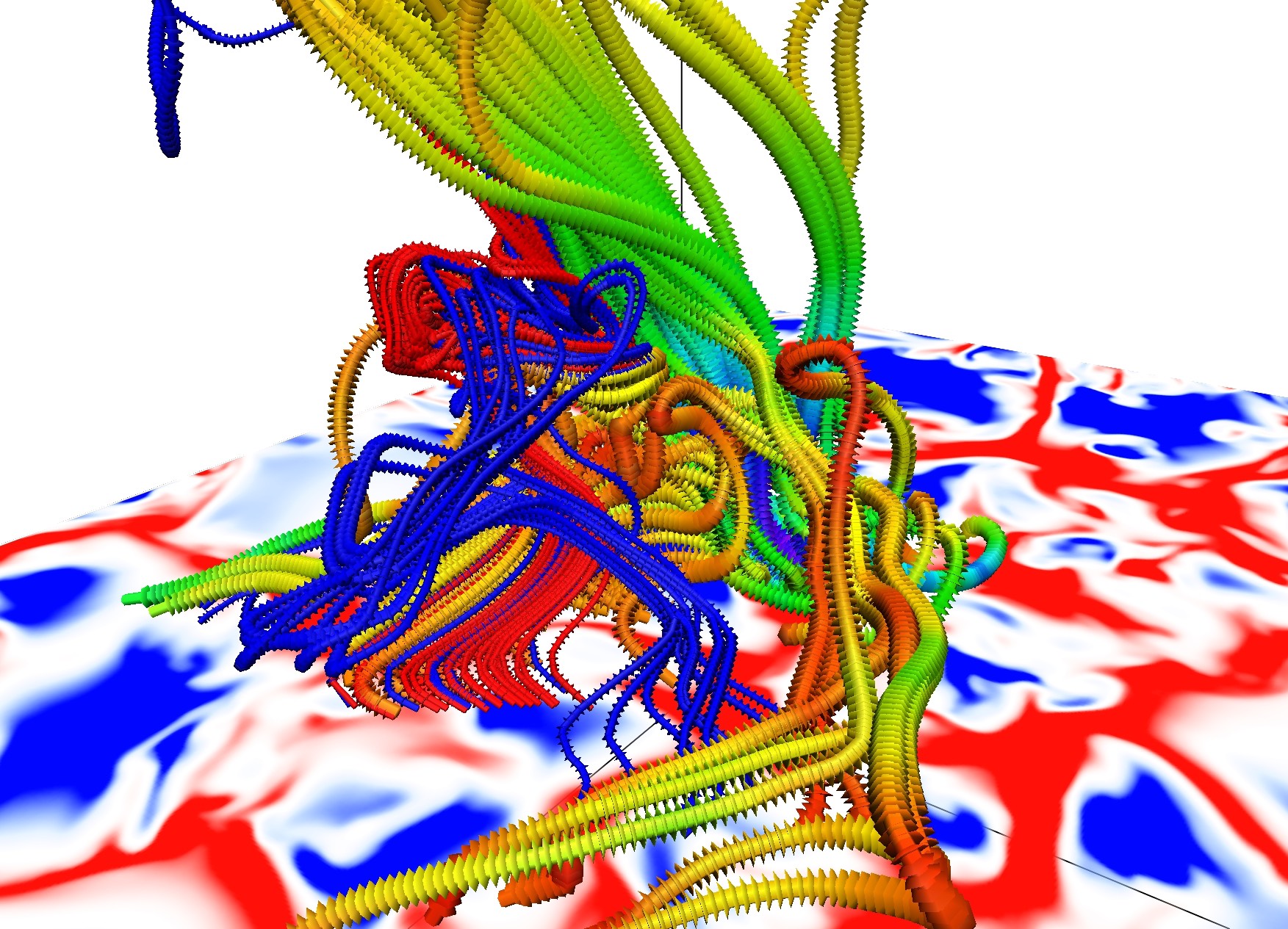}
    \includegraphics[width=0.24\hsize]{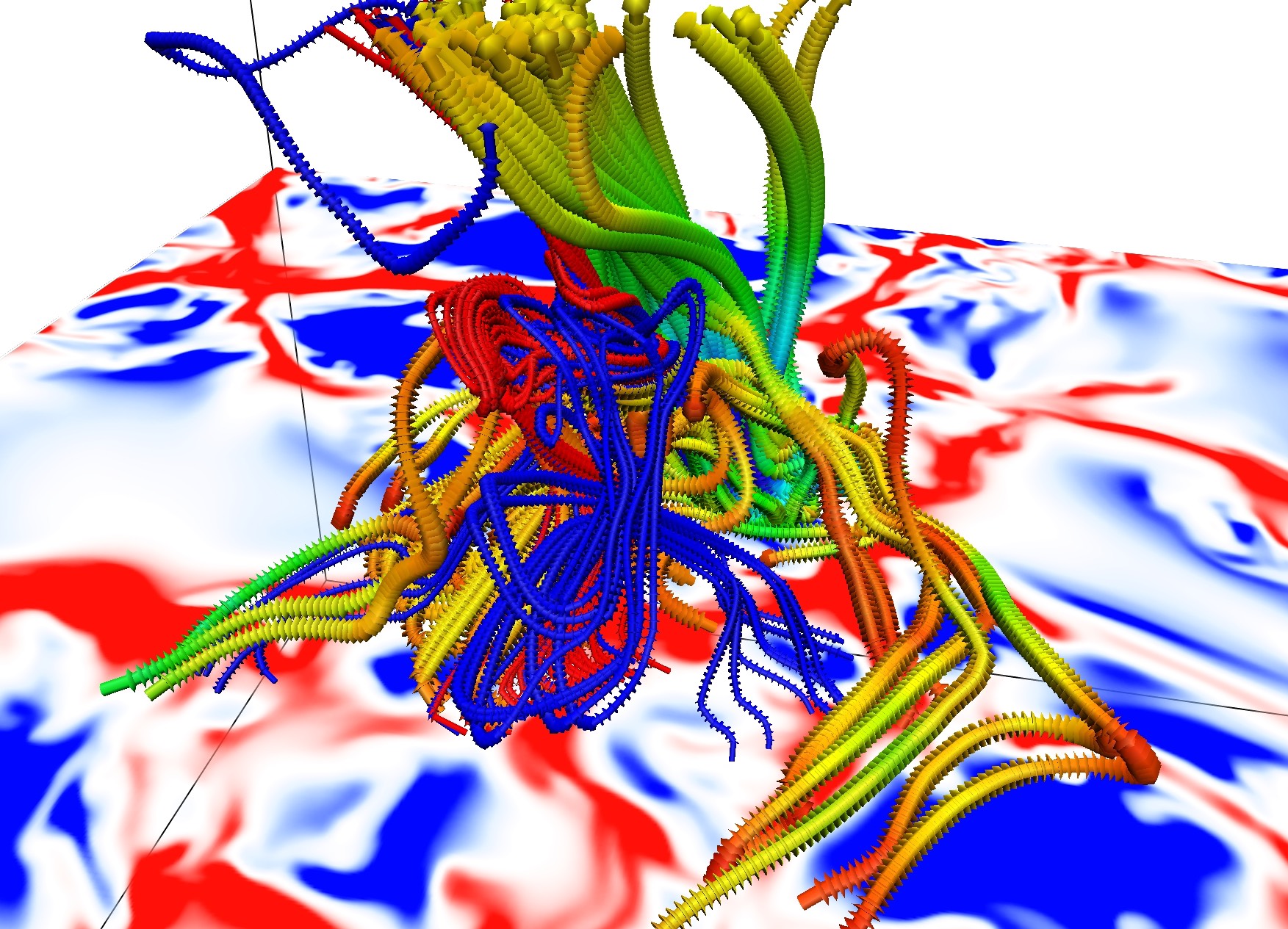}
    \includegraphics[width=0.24\hsize]{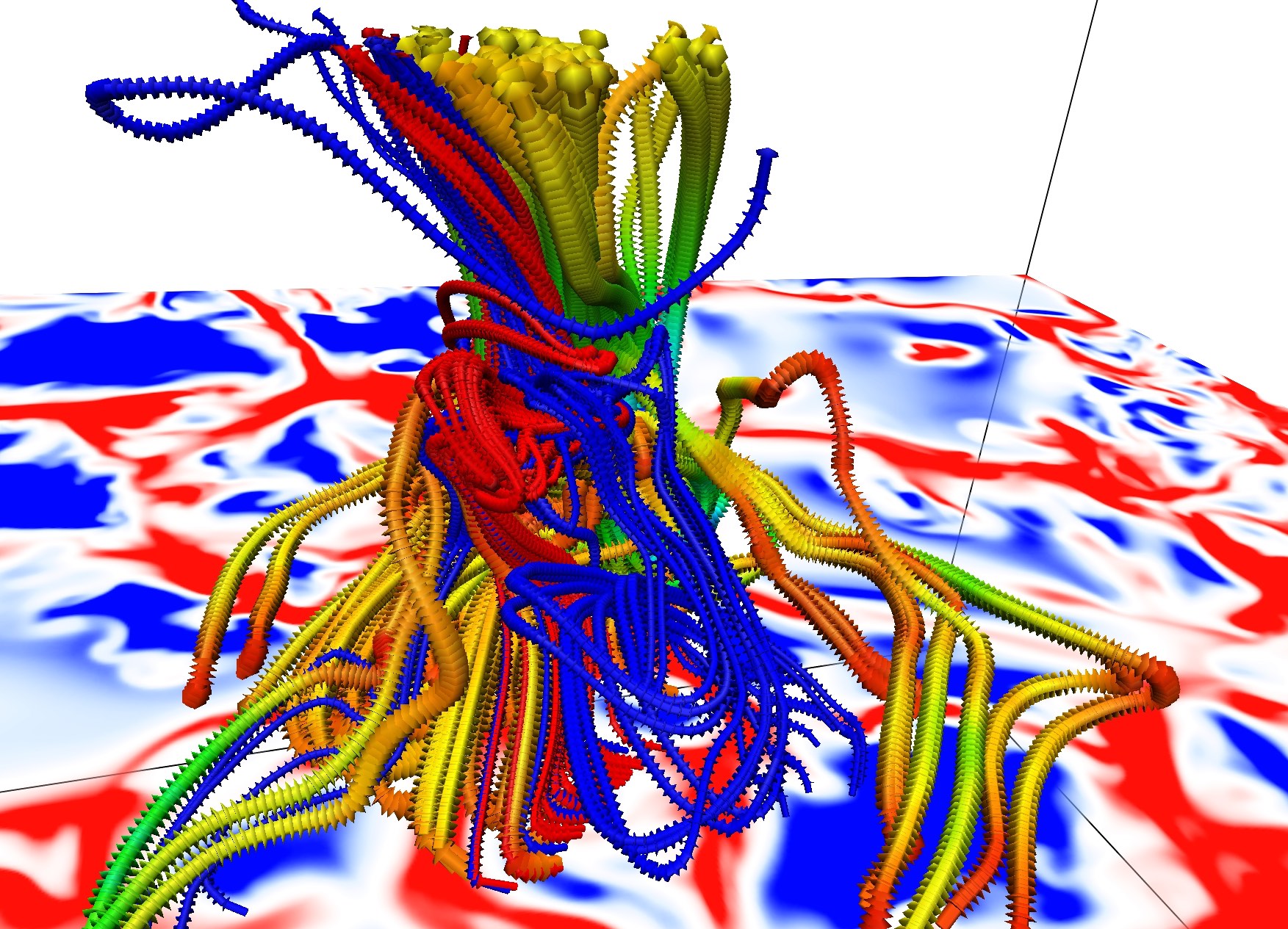}
	\includegraphics[width=0.24\hsize]{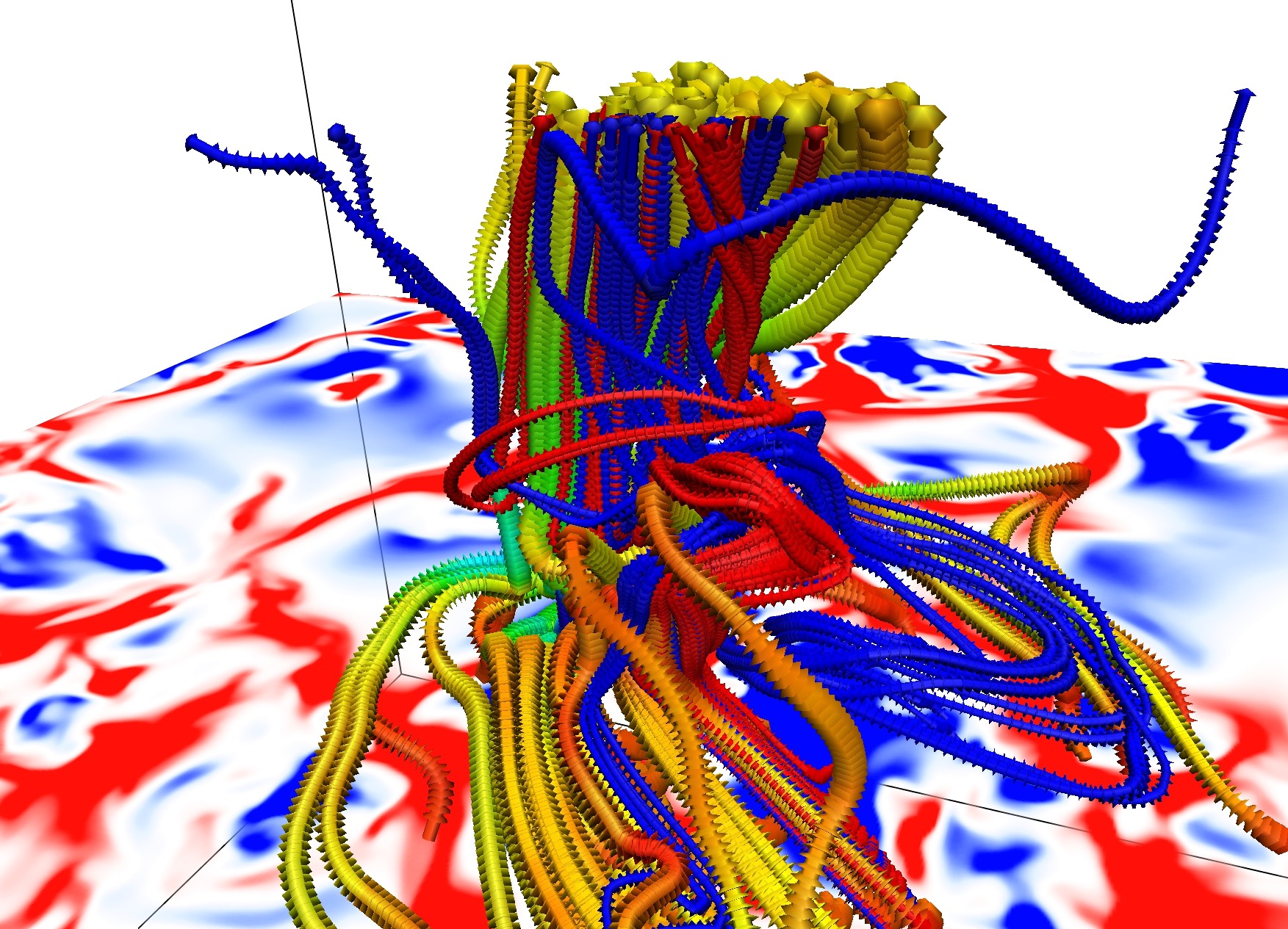}
	\caption{Complex twisted and folded magnetic field (blue and red lines) is reconnected with one of the flux concentrations formed at the chromospheric shock fronts. Different panels correspond to different viewing angles. The blue-white-red horizontal cut shows $\pm2$~km~s$^{-1}$ vertical velocity at $z=0$. See corresponding Movie~6.~\label{fig:chrec}}
\end{figure*}

The complexity of the magnetic field in the chromosphere does not allow us to discern simple magnetic reconnection events. Figure~\ref{fig:chrec} shows an example of reconnection in the chromosphere. This magnetic feature corresponds to the same feature shown in the top panels of Figure~\ref{fig:chst}. The magnetic field lines from the magnetic feature are folded and reconnected
with the same feature within the chromosphere. This happens continuously in the proximity of shock fronts and in the vicinity of the two types of structures described in the previous section. 

\begin{figure}
	\includegraphics[width=0.95\hsize]{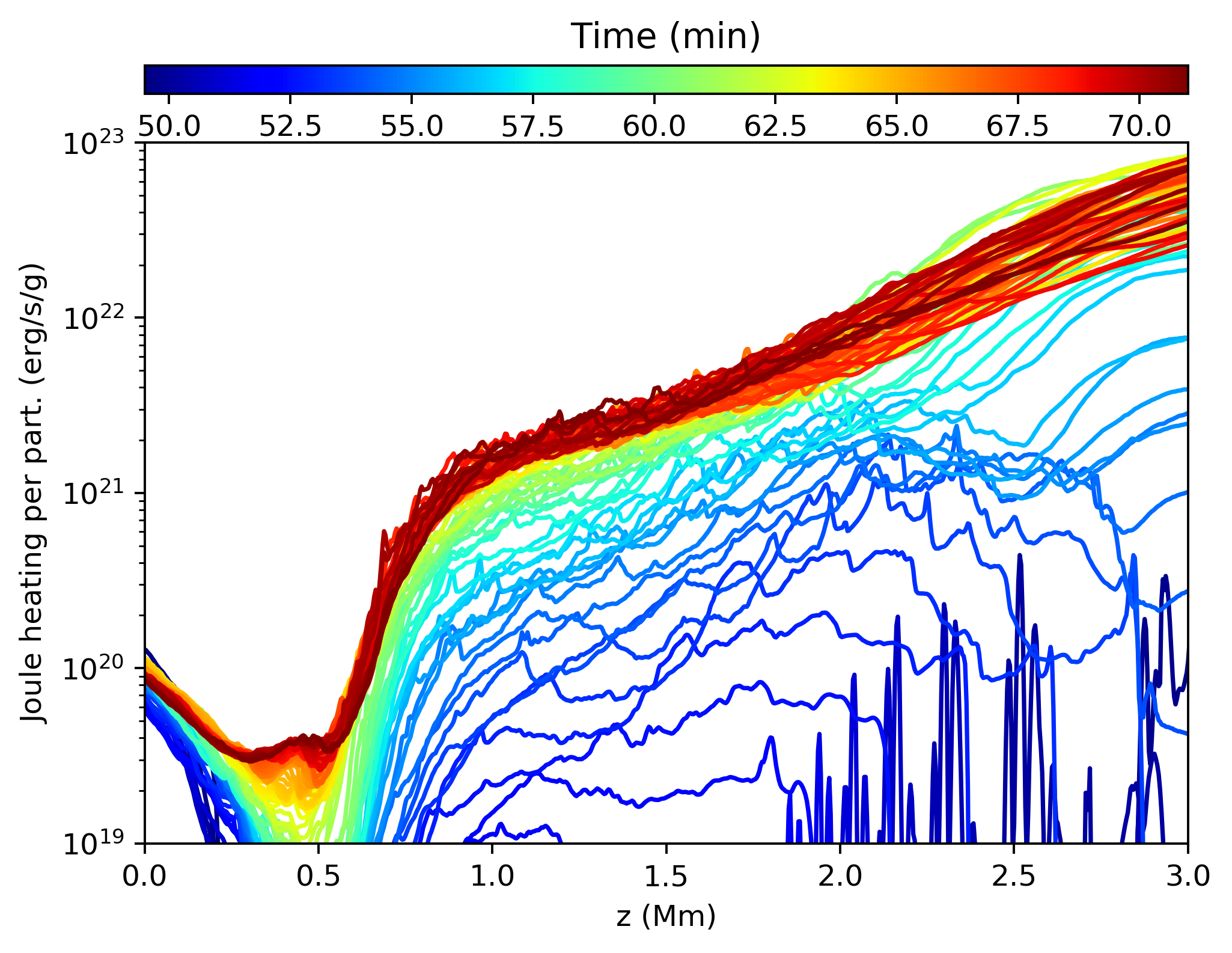}
	\includegraphics[width=0.95\hsize]{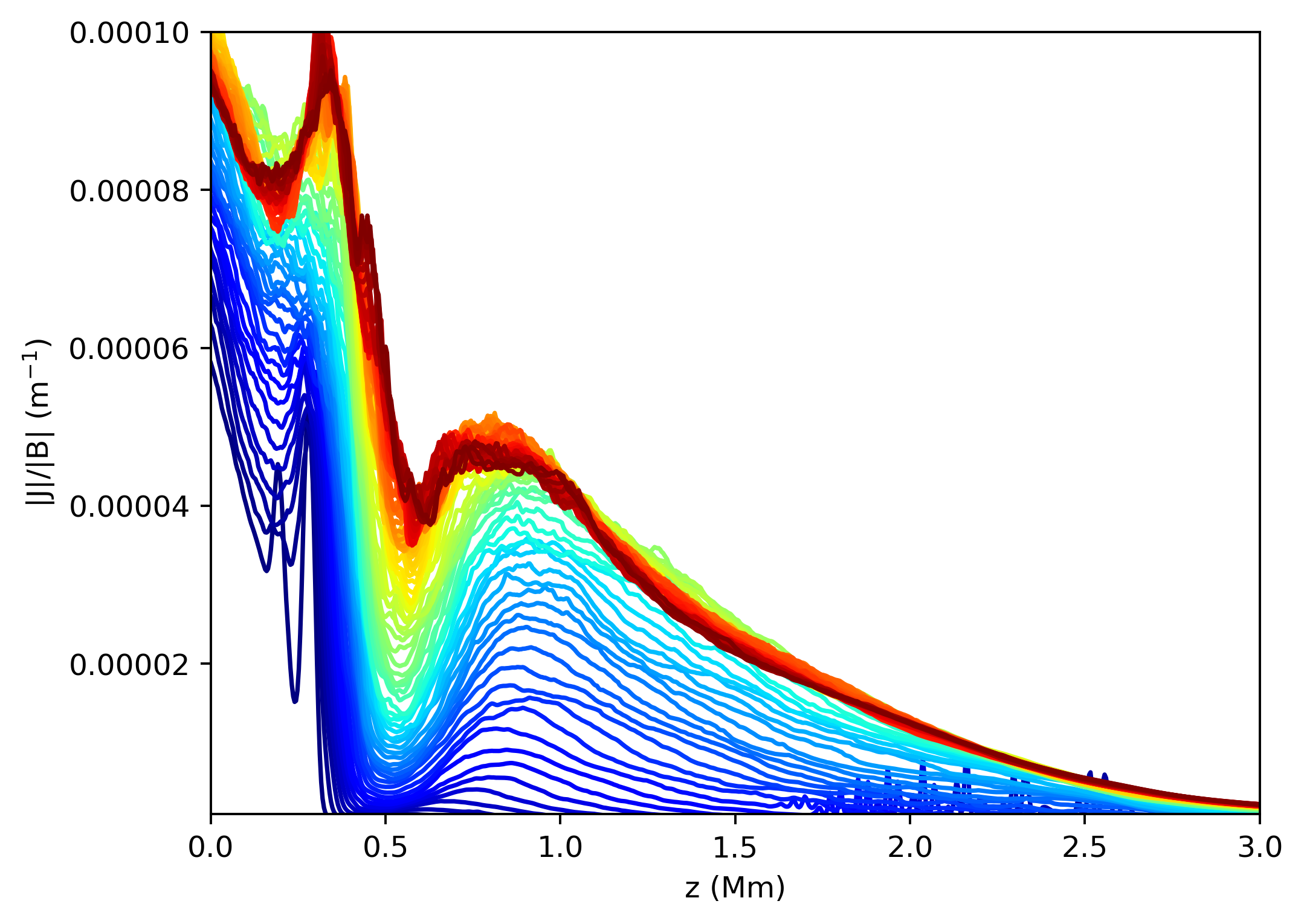}
    \includegraphics[width=0.95\hsize]{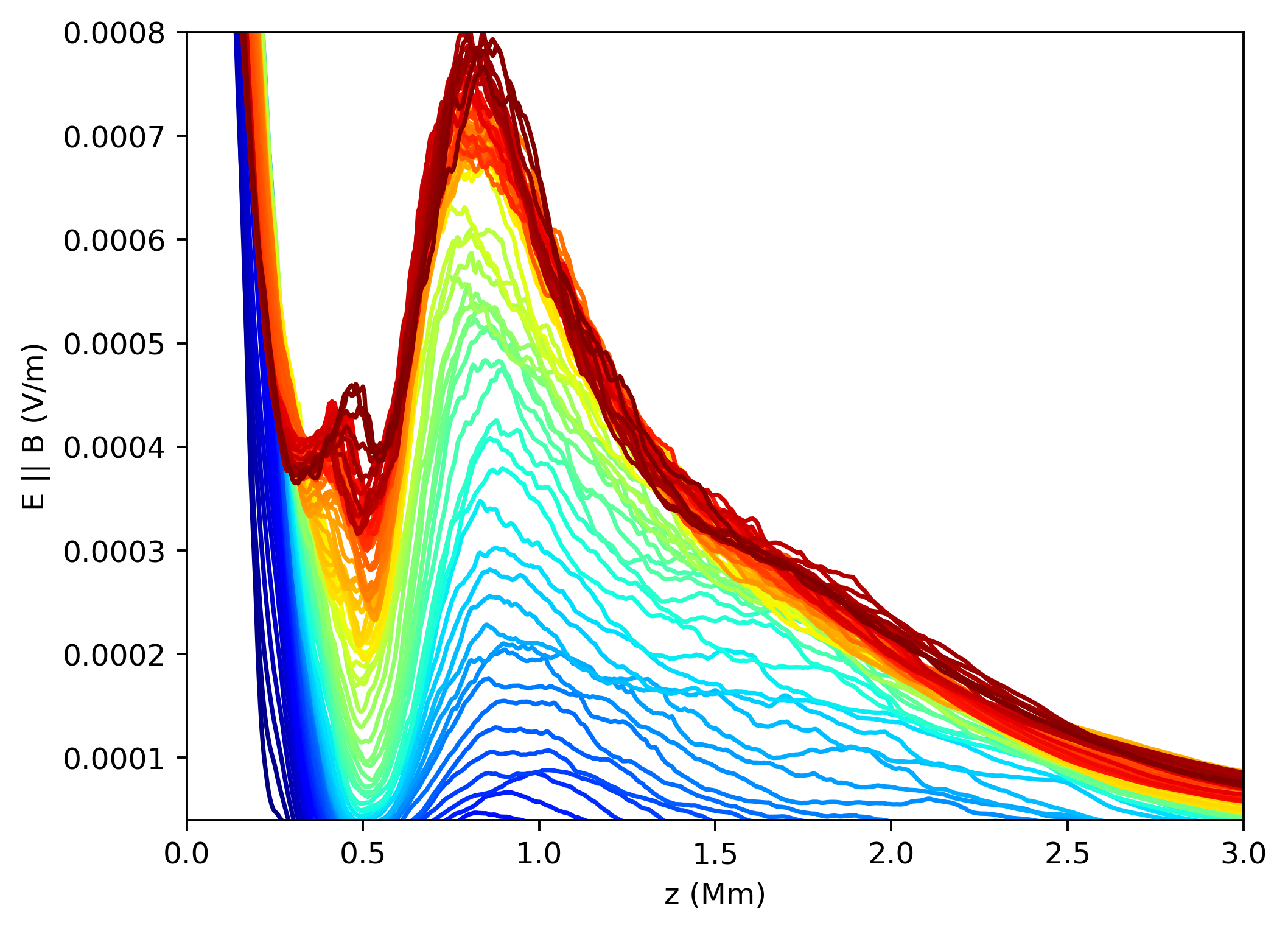}
	\caption{The lower chromosphere reveals larger reconnection processes than the upper photosphere. The horizontal mean of the Joule heating per particle (top) ratio between current and magnetic field strength (middle panel) and electric current parallel to the magnetic field (bottom panel) are shown as a function of height and time (color-scheme). 
    \label{fig:currms}}
\end{figure}

In order to estimate the amplitude of reconnection within the 3D numerical domain we measure the Joule heating per particle, ratio of the current and magnetic strength ($|J|/|B|$) and current parallel to the magnetic field. All these three parameters are indicative of magnetic reconnection and qualitatively provide the same conclusions in our simulation as detailed below. We find that the magnetic field reconnects more often in the lower chromosphere than in its neighboring layers in the simulated atmosphere.
The top panel of Figure~\ref{fig:currms} shows the horizontal mean of the Joule heating per particle as a function of height and time.  The Joule heating per particle drops in the upper photosphere and shows a drastic increase with height in the lower chromosphere. Note the similarities of the Joule heating per particle, kinetic energy per particle (Figure~\ref{fig:brms}) and the various components of the velocity gradient (Figure~\ref{fig:gradheight}) as a function of height within the lower atmosphere. This shows the strong relationship between these three processes in the lower chromosphere. Even though both the current and magnetic field strength decay with height, in the lower chromosphere the ratio of the current and magnetic strength (middle panel of Figure~\ref{fig:currms}) has a local maximum which is indicative of the large number of reconnection events there. It is also remarkable that in the upper photosphere (0.3-0.5~Mm) the heating per particle and $|J|/|B|$ are relatively large compared to neighboring layers. This suggest a higher reconnection rate.
The current parallel to the magnetic field (bottom panel), in agreement with the other two described above, shows lower values in the upper photosphere and a local peak in the lower chromosphere. Consequently, reconnection is a process that leads to a lack of connectivity between the photosphere and chromosphere.

Since the reconnection is in a high plasma $\beta$ regime, i.e., gas pressure dominates, the reconnection processes in the lower chromosphere and upper photosphere do not show reconnection flows. Instead, one can see in Movie~3 that the lower chromosphere is dominated by magneto-acoustic shocks. 
	
\section{Discussion and conclusions}

We have performed a 3D radiative MHD numerical simulation from the upper convection zone to the lower corona. The model has high spatial resolution, a weak initial magnetic field configuration, and does not include the introduction of new magnetic flux through the boundaries. As a result this experiment allows us to study in detail the impact of the photospheric magnetic field from the quiet Sun on the chromosphere as well as the growth of chromospheric magnetic energy in a self-consistent simulated solar atmosphere. However, due to the very specific setup of the simulation and its properties, one must be aware of the rather simplified scenario of this model as follows: the unsigned magnetic field strength in this model resembles a very quiet internetwork region. Further comparisons with photospheric observations are needed to establish how well this model represents the solar atmosphere. Many studies have shown that large-scale connectivity plays a key role in the energy transfer  \citep[e.g.,][]{Gudiksen:2005lr,Hansteen:2007dt,Peter+Gudiksen+Nordlund2004}. Note that our simulation aims to represent a small region of internetwork field. However, and despite the simplified scenario studied here, we can address how a photospheric field impacts the atmosphere and the conversion from kinetic energy into magnetic energy in the chromosphere with internetwork fields. 

The mean magnetic flux from the photosphere can not maintain the amount of magnetic energy in the chromosphere. From previous studies, it is well known that the Poynting flux in the convection zone is negative \citep[e.g.,][]{Nordlund:2008dq}. However, it has been less clear what is happening at greater heights. The treatment of  the radiative transfer in the lower atmosphere is crucial for this \citep[compare our results with those by][which does not include a full treatment of the radiative losses and does not have a self-consistently heated solar atmosphere]{Amari:2015fe}. Our results reveal that the magnetic field generated by kinetic motion acting in the upper convection zone and photosphere does not reach, on average, the chromosphere. It is true that there is some magnetic flux that reaches the chromosphere. However, due to turnover of the convective motions, the super-adiabaticity and strong downflows in the photosphere, the amount of magnetic flux that is removed from the chromosphere into deeper layers is larger than the amount that reaches the chromosphere from below. 

Our numerical experiment shows a mechanism to convert kinetic energy into magnetic energy in the lower chromosphere. This process involves stretch-twist-fold and 
reconnecting the magnetic field, however, it is not a fast-dynamo process since the growth of the magnetic energy is found to be linear and not exponential. The growth is linear because the flow is not fully turbulent. Instead, the chromosphere is dominated by magnetized shocks. Another important property that separates this process from a local dynamo is that it is not a closed system and there is a strong connection of energy and work transfer between different layers. 

The magnetic field strength reached in the lower chromosphere is of the order of a few tens of Gauss. With these values, one would expect that the magnetic features formed due to shocks in the chromosphere could be observed with the new generation of telescopes, i.e., 
the Daniel K. Inouye Solar Telescope \citep[DKIST, ][]{Warner:2018wo} or 
the European Solar Telescope \cite[EST, ][]{Matthews:2016fk}. 
The diagnostic potential must be fully tested with full synthetic Stokes profiles in chromospheric lines like \ion{Ca}{2} H, 8542~\AA , \ion{Na}{1}~D 5896~\AA\ or \ion{Mg}{1} 5173~\AA. These features will be barely connected to deeper layers and highly confined in small regions. 

From these results, a follow up question to address is whether the magnetic energy generated from the kinetic energy could be dissipated into thermal energy. In this case, ion-neutral interaction effects, i.e., ambipolar diffusion \citep{Braginskii:1965ul} and non--equilibrium ionization \citep{Leenaarts:2007sf,Golding:2014fk} should be taken into account self-consistently \citep{Martinez-Sykora:2019hhegol}.
\citet{Khomenko:2018rm} included ambipolar diffusion in local-dynamo simulations that expands into the low chromosphere and showed that the ambipolar diffusion is capable of dissipating the magnetic energy generated by a local-dynamo into thermal energy \citep[see also][]{Khomenko:2012bh,Martinez-Sykora:2017sci,Martinez-Sykora:2017gol}. Missing components in the \citet{Khomenko:2018rm} simulations are a proper treatment of the chromospheric radiative transfer, as well as non-equilibrium ionization. Consequently, their work focuses on the photosphere/upper photosphere. Note that ambipolar diffusion may also impact quantitatively our results/values, but we believe that qualitatively the findings studied in this paper will remain valid. 

\section{Acknowledgments}

We gratefully acknowledge support by NASA grants, NNX16AG90G, NNX17AD33G, and NNG09FA40C (IRIS), NSF grant AST1714955. The simulations have been run on clusters from the Notur project, and the Pleiades cluster through the computing project s1061, s1630, s1980 and s2053 from the High End Computing (HEC) division of NASA. We thankfully acknowledge the support of the Research Council of Norway through grant 230938/F50, through its Center of Excellence scheme, project number 262622,  and through grants of computing time from the Programme for Supercomputing. This work has benefited from discussions at the International Space Science Institute (ISSI) meetings on ``Heating of the magnetized chromosphere'' where many aspects of this paper were discussed with other colleagues. To analyze the data we have used python, Vapor (www.vapor.ucar.edu) and IDL.

\bibliographystyle{aa}
\bibliography{aamnemonic,collectionbib}

\begin{thebibliography}{}

\bibitem[\protect\citeauthoryear{{Abbett}}{{Abbett}}{2007}]{abbett2007}
{Abbett} W.~P., 2007, \apj, 665, 1469, {The Magnetic Connection between the
  Convection Zone and Corona in the Quiet Sun}

\bibitem[\protect\citeauthoryear{{Abbett} \& {Fisher}}{{Abbett} \&
  {Fisher}}{2012}]{Abbett:2012kc}
{Abbett} W.~P.,  {Fisher} G.~H., 2012, \solphys, 277, 3, {Radiative Cooling in
  MHD Models of the Quiet Sun Convection Zone and Corona}

\bibitem[\protect\citeauthoryear{{Acheson}}{{Acheson}}{1979}]{Acheson:1979lr}
{Acheson} D.~J., 1979, \solphys, 62, 23, {Instability by magnetic buoyancy}

\bibitem[\protect\citeauthoryear{{Alexakis}}{{Alexakis}}{2011}]{Alexakis:2011zm}
{Alexakis} A., 2011, \pre, 83, 036301, {Nonlinear dynamos at infinite magnetic
  Prandtl number}

\bibitem[\protect\citeauthoryear{{Amari}, {Luciani}, \& {Aly}}{{Amari} et
  al.}{2015}]{Amari:2015fe}
{Amari} T., {Luciani} J.-F.,  {Aly} J.-J., 2015, \nat, 522, 188, {Small-scale
  dynamo magnetism as the driver for heating the solar atmosphere}

\bibitem[\protect\citeauthoryear{{Archontis} et al.}{{Archontis} et
  al.}{2004}]{archontis2004}
{Archontis} A., {Moreno-Insertis} F., {Galsgaard} K., {Hood} A.,  {O'Shea} E.,
  2004, A\&A, 426, 1047, Emergence of magnetic flux from the convection zone
  into the corona

\bibitem[\protect\citeauthoryear{{Archontis}, {Dorch}, \&
  {Nordlund}}{{Archontis} et al.}{2003}]{Archontis:2003wq}
{Archontis} V., {Dorch} S.~B.~F.,  {Nordlund} {\AA}., 2003, \aap, 397, 393,
  {Numerical simulations of kinematic dynamo action}

\bibitem[\protect\citeauthoryear{{Archontis}, {Dorch}, \&
  {Nordlund}}{{Archontis} et al.}{2007}]{Archontis:2007kk}
{Archontis} V., {Dorch} S.~B.~F.,  {Nordlund} {\AA}., 2007, \aap, 472, 715,
  {Nonlinear MHD dynamo operating at equipartition}

\bibitem[\protect\citeauthoryear{{Biermann}}{{Biermann}}{1950}]{Biermann:1950yg}
{Biermann} L., 1950, Zeitschrift Naturforschung Teil A, 5, 65, {{\"U}ber den
  Ursprung der Magnetfelder auf Sternen und im interstellaren Raum (miteinem
  Anhang von A. Schl{\"u}ter)}

\bibitem[\protect\citeauthoryear{{Braginskii}}{{Braginskii}}{1965}]{Braginskii:1965ul}
{Braginskii} S.~I., 1965, Reviews of Plasma Physics, 1, 205, {Transport
  Processes in a Plasma}

\bibitem[\protect\citeauthoryear{{Brandenburg}}{{Brandenburg}}{2011}]{Brandenburg:2011yk}
{Brandenburg} A., 2011, \apj, 741, 92, {Nonlinear Small-scale Dynamos at Low
  Magnetic Prandtl Numbers}

\bibitem[\protect\citeauthoryear{{Brandenburg}}{{Brandenburg}}{2014}]{Brandenburg:2014mu}
{Brandenburg} A., 2014, \apj, 791, 12, {Magnetic Prandtl Number Dependence of
  the Kinetic-to-magnetic Dissipation Ratio}

\bibitem[\protect\citeauthoryear{{Brandenburg} \& {Subramanian}}{{Brandenburg}
  \& {Subramanian}}{2005}]{Brandenburg:2005pb}
{Brandenburg} A.,  {Subramanian} K., 2005, \physrep, 417, 1, {Astrophysical
  magnetic fields and nonlinear dynamo theory}

\bibitem[\protect\citeauthoryear{{Cameron} \& {Sch{\"u}ssler}}{{Cameron} \&
  {Sch{\"u}ssler}}{2015}]{Cameron:2015rm}
{Cameron} R.,  {Sch{\"u}ssler} M., 2015, Science, 347, 1333, {The crucial role
  of surface magnetic fields for the solar dynamo}

\bibitem[\protect\citeauthoryear{{Carlsson} \& {Leenaarts}}{{Carlsson} \&
  {Leenaarts}}{2012}]{Carlsson:2012uq}
{Carlsson} M.,  {Leenaarts} J., 2012, \aap, 539, A39, {Approximations for
  radiative cooling and heating in the solar chromosphere}

\bibitem[\protect\citeauthoryear{{Cattaneo}}{{Cattaneo}}{1999}]{Cattaneo:1999fr}
{Cattaneo} F., 1999, \apjl, 515, L39, {On the Origin of Magnetic Fields in the
  Quiet Photosphere}

\bibitem[\protect\citeauthoryear{{Centeno} et al.}{{Centeno} et
  al.}{2017}]{Centeno:2017mi}
{Centeno} R., {Blanco Rodr{\'{\i}}guez} J., {Del Toro Iniesta} J.~C., et al.,
  2017, \apjs, 229, 3, {A Tale of Two Emergences: Sunrise II Observations of
  Emergence Sites in a Solar Active Region}

\bibitem[\protect\citeauthoryear{{Childress} \& {Gilbert}}{{Childress} \&
  {Gilbert}}{1995}]{Childress:1995hc}
{Childress} S.,  {Gilbert} A.~D., 1995, {Stretch, Twist, Fold}

\bibitem[\protect\citeauthoryear{{de la Cruz Rodr{\'{\i}}guez} et al.}{{de la
  Cruz Rodr{\'{\i}}guez} et al.}{2013}]{de-la-Cruz-Rodriguez:2013bs}
{de la Cruz Rodr{\'{\i}}guez} J., {De Pontieu} B., {Carlsson} M.,  {Rouppe van
  der Voort} L.~H.~M., 2013, \apjl, 764, L11, {Heating of the Magnetic
  Chromosphere: Observational Constraints from Ca II {$\lambda$}8542 Spectra}

\bibitem[\protect\citeauthoryear{{De Pontieu}}{{De
  Pontieu}}{2002}]{De-Pontieu:2002by}
{De Pontieu} B., 2002, \apj, 569, 474, {High-Resolution Observations of
  Small-Scale Emerging Flux in the Photosphere}

\bibitem[\protect\citeauthoryear{{Finn} \& {Ott}}{{Finn} \&
  {Ott}}{1988}]{Finn:1988jk}
{Finn} J.~M.,  {Ott} E., 1988, Physical Review Letters, 60, 760, {Chaotic flows
  and magnetic dynamos}

\bibitem[\protect\citeauthoryear{{Golding}, {Carlsson}, \&
  {Leenaarts}}{{Golding} et al.}{2014}]{Golding:2014fk}
{Golding} T.~P., {Carlsson} M.,  {Leenaarts} J., 2014, \apj, 784, 30, {Detailed
  and Simplified Nonequilibrium Helium Ionization in the Solar Atmosphere}

\bibitem[\protect\citeauthoryear{{Gudiksen} et al.}{{Gudiksen} et
  al.}{2011}]{Gudiksen:2011qy}
{Gudiksen} B.~V., {Carlsson} M., {Hansteen} V.~H., et al., 2011, \aap, 531,
  A154, {The stellar atmosphere simulation code Bifrost. Code description and
  validation}

\bibitem[\protect\citeauthoryear{{Gudiksen} \& {Nordlund}}{{Gudiksen} \&
  {Nordlund}}{2005}]{Gudiksen:2005lr}
{Gudiksen} B.~V.,  {Nordlund} {\AA}., 2005, \apj, 618, 1020, {An Ab Initio
  Approach to the Solar Coronal Heating Problem}

\bibitem[\protect\citeauthoryear{{Hansteen}, {Carlsson}, \&
  {Gudiksen}}{{Hansteen} et al.}{2007}]{Hansteen:2007dt}
{Hansteen} V.~H., {Carlsson} M.,  {Gudiksen} B., 2007, {3D Numerical Models of
  the Chromosphere, Transition Region, and Corona}, en Astronomical Society of
  the Pacific Conference Series, Vol. 368, {Heinzel} P., {Dorotovi{\v c}} I.,
  {Rutten} R.~J. (eds.), The Physics of Chromospheric Plasmas, p. 107

\bibitem[\protect\citeauthoryear{{Hayek} et al.}{{Hayek} et
  al.}{2010}]{Hayek:2010ac}
{Hayek} W., {Asplund} M., {Carlsson} M., et al., 2010, \aap, 517, A49,
  {Radiative transfer with scattering for domain-decomposed 3D MHD simulations
  of cool stellar atmospheres. Numerical methods and application to the quiet,
  non-magnetic, surface of a solar-type star}

\bibitem[\protect\citeauthoryear{{Hotta}, {Rempel}, \& {Yokoyama}}{{Hotta} et
  al.}{2015}]{Hotta:2015nr}
{Hotta} H., {Rempel} M.,  {Yokoyama} T., 2015, \apj, 803, 42, {Efficient
  Small-scale Dynamo in the Solar Convection Zone}

\bibitem[\protect\citeauthoryear{{Khomenko} \& {Collados}}{{Khomenko} \&
  {Collados}}{2012}]{Khomenko:2012bh}
{Khomenko} E.,  {Collados} M., 2012, \apj, 747, 87, {Heating of the Magnetized
  Solar Chromosphere by Partial Ionization Effects}

\bibitem[\protect\citeauthoryear{{Khomenko} et al.}{{Khomenko} et
  al.}{2017}]{Khomenko:2017sf}
{Khomenko} E., {Vitas} N., {Collados} M.,  {de Vicente} A., 2017, \aap, 604,
  A66, {Numerical simulations of quiet Sun magnetic fields seeded by the
  Biermann battery}

\bibitem[\protect\citeauthoryear{{Khomenko} et al.}{{Khomenko} et
  al.}{2018}]{Khomenko:2018rm}
{Khomenko} E., {Vitas} N., {Collados} M.,  {de Vicente} A., 2018, ArXiv
  e-prints, {Three-dimensional simulations of solar magneto-convection
  including effects of partial ionization}

\bibitem[\protect\citeauthoryear{{Kitiashvili} et al.}{{Kitiashvili} et
  al.}{2015}]{Kitiashvili:2015nr}
{Kitiashvili} I.~N., {Kosovichev} A.~G., {Mansour} N.~N.,  {Wray} A.~A., 2015,
  \apj, 809, 84, {Realistic Modeling of Local Dynamo Processes on the Sun}

\bibitem[\protect\citeauthoryear{{Leenaarts} et al.}{{Leenaarts} et
  al.}{2007}]{Leenaarts:2007sf}
{Leenaarts} J., {Carlsson} M., {Hansteen} V.,  {Rutten} R.~J., 2007, \aap, 473,
  625, {Non-equilibrium hydrogen ionization in 2D simulations of the solar
  atmosphere}

\bibitem[\protect\citeauthoryear{{Mart{\'{\i}}nez Gonz{\'a}lez} \& {Bellot
  Rubio}}{{Mart{\'{\i}}nez Gonz{\'a}lez} \& {Bellot
  Rubio}}{2009}]{Martinez-Gonzalez:2009rp}
{Mart{\'{\i}}nez Gonz{\'a}lez} M.~J.,  {Bellot Rubio} L.~R., 2009, \apj, 700,
  1391, {Emergence of Small-scale Magnetic Loops Through the Quiet Solar
  Atmosphere}

\bibitem[\protect\citeauthoryear{{Mart{\'{\i}}nez Gonz{\'a}lez} et
  al.}{{Mart{\'{\i}}nez Gonz{\'a}lez} et al.}{2010}]{Martinez-Gonzalez:2010kb}
{Mart{\'{\i}}nez Gonz{\'a}lez} M.~J., {Manso Sainz} R., {Asensio Ramos} A.,
  {Bellot Rubio} L.~R., 2010, \apjl, 714, L94, {Small Magnetic Loops Connecting
  the Quiet Surface and the Hot Outer Atmosphere of the Sun}

\bibitem[\protect\citeauthoryear{{Mart{\'{\i}}nez-Sykora} et
  al.}{{Mart{\'{\i}}nez-Sykora} et al.}{2017a}]{Martinez-Sykora:2017gol}
{Mart{\'{\i}}nez-Sykora} J., {De Pontieu} B., {Carlsson} M., et al., 2017a,
  \apj, 847, 36, {Two-dimensional Radiative Magnetohydrodynamic Simulations of
  Partial Ionization in the Chromosphere. II. Dynamics and Energetics of the
  Low Solar Atmosphere}

\bibitem[\protect\citeauthoryear{{Mart{\'{\i}}nez-Sykora} et
  al.}{{Mart{\'{\i}}nez-Sykora} et al.}{2017b}]{Martinez-Sykora:2017sci}
{Mart{\'{\i}}nez-Sykora} J., {De Pontieu} B., {Hansteen} V.~H., et al., 2017b,
  Science, 356, 1269, On the generation of solar spicules and Alfvenic waves

\bibitem[\protect\citeauthoryear{{Mart{\'{\i}}nez-Sykora} et
  al.}{{Mart{\'{\i}}nez-Sykora} et al.}{2019}]{Martinez-Sykora:2019hhegol}
{Mart{\'{\i}}nez-Sykora} J., {Leenaarts} B., J.~{De Pontieu}, {Carlsson} M., et
  al., 2019, \apj, 847, {Modeling of ion-neutral interaction effects and
  non-equilibrium ionization in the solar chromosphere}

\bibitem[\protect\citeauthoryear{{Matthews} et al.}{{Matthews} et
  al.}{2016}]{Matthews:2016fk}
{Matthews} S.~A., {Collados} M., {Mathioudakis} M.,  {Erdelyi} R., 2016, {The
  European Solar Telescope (EST)}, en \procspie, Vol. 9908, Ground-based and
  Airborne Instrumentation for Astronomy VI, p. 990809

\bibitem[\protect\citeauthoryear{{Moreno-Insertis} et al.}{{Moreno-Insertis} et
  al.}{2018}]{Moreno-Insertis:2018hl}
{Moreno-Insertis} F., {Martinez-Sykora} J., {Hansteen} V.~H.,  {Mu{\~n}oz} D.,
  2018, \apjl, 859, L26, {Small-scale Magnetic Flux Emergence in the Quiet Sun}

\bibitem[\protect\citeauthoryear{{Nordlund}}{{Nordlund}}{2008}]{Nordlund:2008dq}
{Nordlund} {\AA}., 2008, Physica Scripta Volume T, 133, 014002, {Stellar
  (magneto-)convection}

\bibitem[\protect\citeauthoryear{{November} \& {Simon}}{{November} \&
  {Simon}}{1988}]{November:1988oj}
{November} L.~J.,  {Simon} G.~W., 1988, \apj, 333, 427, {Precise proper-motion
  measurement of solar granulation}

\bibitem[\protect\citeauthoryear{{Parker}}{{Parker}}{1955}]{Parker:1955rc}
{Parker} E.~N., 1955, \apj, 122, 293, {Hydromagnetic Dynamo Models.}

\bibitem[\protect\citeauthoryear{{Peter}, {Gudiksen}, \& {Nordlund}}{{Peter} et
  al.}{2004}]{Peter+Gudiksen+Nordlund2004}
{Peter} H., {Gudiksen} B.~V.,  {Nordlund} {\AA}., 2004, Apjl, 617, L85,
  {Coronal Heating through Braiding of Magnetic Field Lines}

\bibitem[\protect\citeauthoryear{{Rempel}}{{Rempel}}{2014}]{Rempel:2014sf}
{Rempel} M., 2014, \apj, 789, 132, {Numerical Simulations of Quiet Sun
  Magnetism: On the Contribution from a Small-scale Dynamo}

\bibitem[\protect\citeauthoryear{{Rempel}}{{Rempel}}{2017}]{Rempel:2017zl}
{Rempel} M., 2017, \apj, 834, 10, {Extension of the MURaM Radiative MHD Code
  for Coronal Simulations}

\bibitem[\protect\citeauthoryear{{Sainz Dalda} et al.}{{Sainz Dalda} et
  al.}{2012}]{Sainz-Dalda:2012qf}
{Sainz Dalda} A., {Mart{\'{\i}}nez-Sykora} J., {Bellot Rubio} L.,  {Title} A.,
  2012, \apj, 748, 38, {Study of Single-lobed Circular Polarization Profiles in
  the Quiet Sun}

\bibitem[\protect\citeauthoryear{{Schekochihin} et al.}{{Schekochihin} et
  al.}{2004}]{Schekochihin:2004tg}
{Schekochihin} A.~A., {Cowley} S.~C., {Taylor} S.~F., {Maron} J.~L.,
  {McWilliams} J.~C., 2004, \apj, 612, 276, {Simulations of the Small-Scale
  Turbulent Dynamo}

\bibitem[\protect\citeauthoryear{{Skartlien}}{{Skartlien}}{2000}]{Skartlien2000}
{Skartlien} R., 2000, \apj, 536, 465, {A Multigroup Method for Radiation with
  Scattering in Three-Dimensional Hydrodynamic Simulations}

\bibitem[\protect\citeauthoryear{{V{\"o}gler} \& {Sch{\"u}ssler}}{{V{\"o}gler}
  \& {Sch{\"u}ssler}}{2007}]{Vogler:2007yg}
{V{\"o}gler} A.,  {Sch{\"u}ssler} M., 2007, \aap, 465, L43, {A solar surface
  dynamo}

\bibitem[\protect\citeauthoryear{{Warner} et al.}{{Warner} et
  al.}{2018}]{Warner:2018wo}
{Warner} M., {Rimmele} T.~R., {Pillet} V.~M., et al., 2018, {Construction
  update of the Daniel K. Inouye Solar Telescope project}, en Society of
  Photo-Optical Instrumentation Engineers (SPIE) Conference Series, Vol. 10700,
  Society of Photo-Optical Instrumentation Engineers (SPIE) Conference Series,
  p. 107000V

\end{thebibliography}

\end{document}